\documentclass{aa}  
\usepackage[most]{tcolorbox}
\usepackage[export]{adjustbox}
\usepackage{tcolorbox}
\usepackage{xcolor} % for colors

\usepackage{graphicx}
\usepackage{txfonts}
\usepackage{hyperref}
\usepackage{lipsum}
\usepackage[dvipsnames]{xcolor}
\usepackage{subcaption}
\usepackage{nicematrix}
\usepackage{makecell} \setcellgapes{2pt}
\hypersetup{
colorlinks=true,
linkcolor=RubineRed,
filecolor=NavyBlue,
urlcolor=NavyBlue,
citecolor=NavyBlue
}
\begin{document} 

   \title{A MUSE Study of Stellar and Ionized-Gas Kinematics in Central Group Galaxies}
\titlerunning{Stellar and Ionized-Gas in Central Group Galaxies}

   \author{V. Olivares\inst{1,}\inst{2}, E. O'Sullivan\inst{3}, P. Temi\inst{4}, S. Suárez\inst{1}, K. Kolokythas\inst{5,6,7}, M. Gaspari\inst{8}, F. Ubertosi\inst{9,10}, M. Aravena\inst{11,12}, F. Combes\inst{13}, Y. Su\inst{14}, P. Salomé\inst{13}, E. Treister\inst{15}, M. Gitti\inst{16}, F. Brighenti\inst{17}, and M.-L. Gendron-Marsolais\inst{18}
   }
   
\authorrunning{Olivares et al.}
   \institute{
    Departamento de F\'isica, Universidad de Santiago de Chile, Av. Victor Jara 3659, Santiago 9170124, Chile \email{valeria.olivares@usach.cl}
    \and % 1 
    Center for Interdisciplinary Research in Astrophysics and Space Exploration (CIRAS), Universidad de Santiago de Chile, Santiago 9170124, Chile % 2
    \and
    Center for Astrophysics $|$ Harvard \& Smithsonian, 60 Garden Street, Cambridge, MA, 02138, USA % 3 EO
    \and
    Astrophysics Division, NASA—Ames Research Center, MS 245-6, Moffett Field, CA 94035, USA % PT
    \and
    Centre for Radio Astronomy Techniques and Technologies, Department of Physics and Electronics, Rhodes University, Makhanda 6140, South Africa
    \and South African Radio Astronomy Observatory, 2 Fir Street, Observatory 7925, South Africa
    \and INAF -- Istituto di Radioastronomia, via Gobetti 101, I–40129 Bologna, Italy 
    \and
    Department of Physics, Informatics and Mathematics, University of Modena and Reggio Emilia, 41125 Modena, Italy % MG
    \and
    Dipartimento di Fisica e Astronomia, Università di Bologna, via Gobetti 93/2, I-40129 Bologna, Italy % FU
    \and
    Istituto Nazionale di Astrofisica - Istituto di Radioastronomia (IRA), via Gobetti 101, I-40129 Bologna, Italy %FU
    \and
    Instituto de Estudios Astrofícos, Facultad de Ingeniería y Ciencias, Universidad Diego Portales, Av. Ejército 441, Santiago, Chile % MA
    \and
    Millenium Nucleus for Galaxies (MINGAL) % MA 
    \and
    LUX, Observatoire de Paris, Collège de France, PSL University, CNRS, Sorbonne University, Paris, France % FC 
    \and
    Department of Physics and Astronomy, University of Kentucky, 505 Rose Street, Lexington, KY 40506, USA % YS
    }

   \date{Received XX; accepted YY}

 \abstract{
    Understanding the gas accretion and feedback processes in low-mass galaxy groups is crucial for bridging the gap between isolated ellipticals and cluster-dominant galaxies. We present new Multi Unit Spectroscopic Explorer (MUSE) observations of 17 nearby Brightest Group Galaxies (BGGs) that reveal a complex and multiphase origin for their gas reservoirs. Warm ionized gas traced with H$\alpha$ and [NII] emission lines is detected in 77\% (13/17) of the sample, displaying diverse morphologies from rotating disks with/without tails and diffuse gas, rings, and filamentary structures. While most of our BGGs host regularly rotating stellar components, unlike Brightest Cluster Galaxies (BCGs), their warm ionized gas exhibits a wide range of kinematic. Nearly half of the rotating gas disks are aligned with the stars, consistent with internal recycling. However, the presence of tails, diffuse gas, or rings in three of these systems suggests that these structures may retain signatures of past accretion or interactions despite their present-day kinematic alignment. The remaining disks, ring-like structures, and most filamentary systems are misaligned, supporting an external origin through mergers, interactions, or cooling from their hot halos. Compact disks are consistent with centrally-condensed, low-angular-momentum gas, possibly linked to residual cooling. Most filamentary systems lie near the precipitation threshold ($t_{\rm cool}/t_{\rm eddy}\sim 1$--2) and preferentially occupy the turbulence-dominated locus of the k-plot, supporting a condensation-driven origin consistent with Chaotic Cold Accretion.
    All galaxies with detected X-ray halos, including those without warm ionized gas, exhibit low central entropy ($\rm <30\, keV\, cm^{2}$) and short central cooling times ($\rm <3\, Gyr$), indicating that these thermodynamic conditions alone are not sufficient to predict the presence of multiphase gas, and that additional processes, such as AGN- or interaction-driven perturbations and turbulence-induced thermal instabilities, play a key role in regulating gas condensation. Together, these results indicate that no single mechanism dominates, rather, BGGs are transitional systems where internal recycling, environmental accretion, and cooling-driven condensation coexist, linking isolated ellipticals and cluster-dominant galaxies. 
    }

   \maketitle

\section{Introduction}

Contrary to the stereotypical impression of ``red and dead,’’ many early-type galaxies (ETGs) contain reservoirs of multiphase gas and show emission lines. Their origin is crucial for understanding their quenching pathways and informing their potential rejuvenation. Much observational effort has been devoted to cluster-dominant galaxies (so-called Brightest Cluster Galaxies, or BCGs), particularly using Integral Field Unit (IFU) observations, which have enabled spatially resolved studies of their gas kinematics, ionization mechanisms, and metallicities \citep[e.g.,][]{salome11,Edge2001,2019ApJ...870...57V,2012ApJ...744..112L,hamer2019,Olivares2019,russell19,olivares23,Poitras2026}. It is now widely accepted that multiphase gas in BCGs primarily forms through cooling of the intracluster medium (ICM). In contrast, nebular emission in isolated early-type galaxies may have a more diverse origin, including stellar mass loss, galaxy interactions,. and cooling \cite{Eskenasy2024,2026OJAp....958379E}. Overall, group-dominant galaxies or BGGs remain relatively unstudied as a population compared to both BCGs and the general population of elliptical galaxies, despite the fact that galaxy groups represent the environment of most nearby galaxies and are the building blocks of massive clusters.

Several galaxy groups exhibit remarkable similarities to clusters, hosting a hot intragroup medium (IGrM; e.g. \citealt{dong10, osullivan17}) that cools via X-ray emission in a manner analogous to the ICM. However, groups are inherently more diverse, and their lower velocity dispersions and smaller galaxy separations make mergers more frequent than in clusters. Also ram-pressure stripping is less effective than in clusters, thus galaxies are able to retain more of their cold gas before falling into the group core. All of these factors allows a larger fraction of the cold gas from infalling gas-rich spirals to reach the central regions, influencing the overall gas content. Moreover, cold molecular gas appears to be more common in group-dominant galaxies than in other ellipticals \citep{osullivan15}. AGN feedback is expected to have a larger impact on groups than on clusters due to their shallower potential wells \cite{Giodini2009}. Overall the the interplay between AGN and environmental processes may be important in shaping the evolution of the most massive galaxies \citep{Vulcani2026}.

Central elliptical galaxies in massive groups shows extended filamentary warm ionized gas has been detected along X-ray cavities in several systems \citep{Temi2018, Temi2022, Werner2019, schellenberger20, Osullivan2024, Pasini2021, Temi2026}. These cavities are typically inflated by radio jets, providing clear evidence of ongoing mechanical AGN feedback. \citet{Ubertosi2025} further show that current AGN activity is strongly linked to H$\alpha$ emission peaking at the nucleus, suggesting that the presence and morphology of warm ionized gas may serve as an important tracer of both past and present feedback episodes. 

To investigate how galaxies in groups acquire, retain and process their gas, the Complete Local-Volume Groups Sample (CLoGS; \citealt{osullivan17}) was constructed. CLoGS represents the first statistically complete, optically selected, and representative sample of local (D $<$ 80 Mpc) galaxy groups observed across X-ray (O'Sullivan et al. in prep, \citealt{osullivan17}), multi-frequency radio \citep{kolokythas18, Kolokythas2019}, mid-IR, FUV \citep{Kolokythas2022}, and optical wavelengths. The sample comprises 53 groups with Richness $R$\footnote{Richness (R) is defined as the number of galaxies within the group with B-band luminosity log L$_{\rm B}$ $>$ 10.2 \citep{osullivan17}.} between 2 and 8. The molecular gas content of all CLoGS groups has been studied by single-dish observations \citep{osullivan15, osullivan18b}.

Multi Unit Spectroscopic Explorer (MUSE, \citealt{Bacon2010}) observations of the central galaxy of the high-richness ($R>3$) CLoGS sub-sample reveal optical emission lines in 17 out of 18 systems, with filamentary structures observed in most systems \citep{Olivares2022}. Notably, a few sources host unusually large gas disks and rings that extend over several kpc. Such extended, coherent structures are rare in BCGs, which more commonly exhibit filamentary nebulae. Kinematic studies of the stellar component of such systems reveal that about half are fast rotators, showing aligned morpho-kinematic axes and oblate shapes, while the remainder are slow rotators spanning oblate and prolate morphologies. These differences correlate with their multiphase gas and feedback properties: fast rotators typically retain cold gas acquired through gas-rich mergers or interactions, whereas slow rotators generally reflect gas-poor merger histories, with any detected cold gas likely originating from cooling of their hot halos \citep{Loubser2022}. 

A key modern framework to interpret the interplay between multiphase gas, turbulence, and AGN self-regulation is Chaotic Cold Accretion (CCA; \citealt{gaspari13,gaspari17,gaspari20}, for a review). In this picture, radiative cooling in hot halos competes with turbulent eddies and promotes the nonlinear condensation of cold clouds and warm filaments when the local cooling time approaches the eddy turnover time. The ensuing multiphase rain intermittently fuels the central engine and sustains a self-regulated feedback loop, naturally producing extended, clumpy filamentary structures and disturbed kinematics. This framework is particularly relevant for low-richness groups, where cooling, angular momentum, mergers, and AGN-driven stirring may all contribute comparably, potentially bridging the more settled gas structures often found in ellipticals and the filament-dominated condensation seen in cluster cores. It therefore motivates observational tests that combine thermodynamic and kinematic information, such as the k-plot and the $C \equiv t_{\rm cool}/t_{\rm eddy}$ criterion (\citealt{gaspari18,juranova20,Maccagni2021,Olivares2022,Mehdipour2023}), which we explore below.

This paper aims to investigate the origin of the warm ionized gas in the central galaxies of low-richness groups drawn from the CLoGS survey, using new MUSE observations. This work extends our previous MUSE study of high-richness systems to lower-richness environments 
\citep{Olivares2022,Lagos2022,Loubser2022}, and describe the morphology and kinematics of the both gas and stellar component. We leave the ionization analysis and a full comparison between both subsamples for future work. The paper is organized as follows: Section \ref{sec:sample} describes the sample, Section \ref{sec:muse} details the observations used throughout this study, the results are presented in Section \ref{sec:results}, and Section \ref{sec:discussion} provides a discussion of the findings.

\section{Sample}\label{sec:sample}
\begin{figure}[ht!]
    \centering
    \includegraphics[scale=0.8]{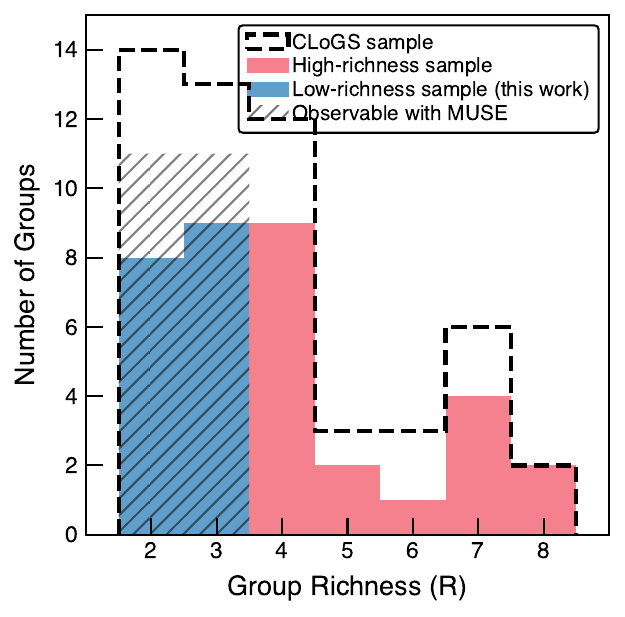}
    \caption{CLoGS samples. The black line shows the full CLoGS sample, the pink outline corresponds to the high-richness subsample previously analyzed with MUSE and Chandra \citep{Olivares2022}, and the blue outline marks the low-richness subsample presented in this work. We have also highlighted the galaxies observables with MUSE ($\rm dec<35\deg$) with diagonal lines.}
    \label{fig:sample_distribution}
\end{figure}

To study the properties and origin of the optical line-emitting gas in nearby group-dominant galaxies, we obtained MUSE observations of 17 sources selected from the low-richness subsample of CLoGS (R = 2–3), which contains a total of 27 sources \citep{osullivan15, osullivan17, osullivan18b}). Our 17 targets represent the majority of the 22 CLoGS low-richness systems observable with MUSE from the Southern Hemisphere (declination $\lesssim 35^\circ$). All systems were originally proposed, but only 17 were observed.

The full CLoGS sample comprises 53 galaxy groups, identified in the Lyon-Meudon Extragalactic Database (LEDA, \citealt{makarov14}) using friends-of-friends and hierarchical clustering algorithms. Each group was required to contain an elliptical galaxy as its dominant member. This optically selected approach makes CLoGS an ideal dataset for investigating the origin of the gas in lower-mass systems. A detailed discussion of sample selection is given in \citet{osullivan17}. Figure \ref{fig:sample_distribution} shows the richness distribution of the entire CLoGS sample (black line), highlighting the high-richness ($R>3-8$) subsample previously studied with MUSE and {\it Chandra} \citep{Olivares2022}, and the low-richness subsample (R=2-3) analyzed in this work.

Our MUSE sample spans a wide range of stellar masses, X-ray and radio properties. The stellar masses span a factor of approximately 2-3, from 0.5$\times$10$^{11}$ to 1.2$\times$10$^{11}$~M$_{\odot}$. The stellar masses were determined using a mass-to-light ratio model, and using galaxy colors from SDSS \textit{ugriz} and 2MASS (2 Micron All Sky Survey) \textit{JHK} magnitudes \citep{osullivan18b}. 

Radio jets have been detected in 2 of the 17 sources, covering physical scales of 16 and 244 kpc, based on 1.4 GHz Very Large Array (VLA) and Giant Metrewave Radio Telescope (GMRT) observations at 610 MHz and 235 MHz \citep{Kolokythas2019}. In 7/17 sources the radio emission is instead very compact and point-like in nature ($\sim5.5\arcsec \times4.0\arcsec$), while 2/17 show diffuse or a radio-remnant emission are based on the high 235 MHz GMRT resolution \citep{Kolokythas2019}. In \citet{Kolokythas2019} they reported the contribution from star formation to the radio emission. Based on that analysis, NGC 128, NGC 1395, NGC 1779, NGC 1395, NGC 3923, NGC 4697, and NGC 5061, NGC 5903, NGC 7252, lie close to the relation expected for radio emission powered by star formation, as electrons are accelerated during the supernovae phase, giving rise to synchrotron emission, suggesting that their radio emission may be dominated by, or have a significant contribution (SF contribution is $>$50\%-60\%). NGC 4697 is identified as star formation dominated. In contrast, NGC 2911 is dominated by AGN activity. In NGC 5903, the most likely origin for the radio emission is a combination of interaction-triggered AGN outburst, and high-velocity collisions between galaxies \citep{Kolokythas2019}.

With respect to their X-ray properties, 9 out of 17 systems host group-scale halos based on \textit{XMM-Newton} and \textit{Chandra} observations (O'Sullivan et al., in prep.), while 5 out of 17 exhibit galaxy-scale halos. The remaining 3 out of 17 systems show no detected X-ray halo. The MUSE observations and the key multiwavelength properties of each target are summarized in Table~\ref{tab:observations} and Table~\ref{tab:sample_prop}, respectively.

\section{MUSE observations and analysis}\label{sec:muse}

\subsection{Observations}

The central group galaxies were observed with the MUSE instrument on the ESO VLT (program 0110.A-4397(A) and 0112.A-2375(A); PI: V. Olivares) between cycles 110 and 112. We also analyzed the MUSE observations of the group-dominant galaxy: NGC 7252 (099.B-0281(A), PI: G. Privon). MUSE is a wide-field optical integral field spectrograph covering $\sim$0.46–0.93 $\mu$m, with a $1\arcmin \times 1\arcmin$ field of view and a spatial sampling of $0.2\arcsec$ in wide-field mode (WFM). The spectral resolution is $R \sim 3000$ at the central wavelength. 

The observations consisted of four pointings for each target: three exposures of 870–900s centered on each galaxy group, and a fourth 120s exposure on a nearby blank field to obtain an uncontaminated sky spectrum. For the case of NGC 5629, the galaxy was only observed for 900s. The MUSE observations of NGC 7252 consist of 4 pointings with slight offsets around the center of the galaxy, and with a total exposure time of 5h and 34s. The seeing conditions during the observations ranged from $0.5\arcsec$ to $1.9\arcsec$. Further details for each source are provided in Table~\ref{tab:observations}.

\subsection{Analysis}
The MUSE observations were corrected for Galactic extinction using the \citet{odonnell94} extinction law in combination with the dust maps of \citet{schlegel98}. The astrometry of our sources was refined by matching stars and galaxies in the field of view with r-band images from the Pan-STARRS1 Image Access database.

To determine the systemic velocity of each galaxy, we used a Python implementation of Penalized PiXel Fixing ({\sc pPXF}) following the procedure outlined in \citet{2017MNRAS.466..798C}. To determine system velocity of each target, a spectrum was extracted from the central 0.5 arcsecond aperture and fitted using the E-MILES libraries to model the stellar continuum \citep{Vazdekis2016}, with all emission lines masked. The redshift and corresponding systemic velocity were derived using the relation $v_{\rm sys} = cz$, 
and for small velocity differences in the LOSVD $\Delta$V = V = c\,$\ln(1+z)$ \citep{2017MNRAS.466..798C}. All gaseous and stellar velocities reported in this work are expressed relative to this systemic rest frame. The redshifts obtained through this method are listed in Table~\ref{tab:observations}.

To subtract the stellar and gas distribution and kinematics, we modeled the stellar continuum and nebular emission simultaneously using \texttt{pPXF} for each spaxel \citep{Cappellari2011}. We first produced {\sc Voronoi-binned} maps of the stellar continuum with a target signal-to-noise ratio of 80. The stellar continuum in each bin was then modeled using the E-MILES stellar library \citep{Vazdekis2016}, masking all emission lines during the fitting process. The resulting best-fitting (optimal stellar template) linear combination of stellar templates, including the stellar kinematics derived for each bin, was then used as the stellar continuum model for all individual spaxels belonging to that Voronoi bin. For each spaxel, the normalization of the stellar template was allowed to vary to account for the local continuum level, while the stellar population mixture and kinematics were fixed to those determined from the corresponding Voronoi bin. 

Later, the optical emission lines were fitted together on a spaxel-by-spaxel basis. Emission lines were modeled using single Gaussian components, with velocities and velocity dispersions tied. We also tested allowing independent kinematics for the Balmer and forbidden lines, yielding consistent results. Our procedure follows the general strategy of spatially resolved spectral fitting implemented in the GIST pipeline \citep[e.g.][]{Bittner2019}.

In regions showing evidence for broad or kinematically complex emission, we additionally tested a two-component Gaussian decomposition, and adopted only when it resulted in an improved $\chi2$; otherwise, the single Gaussian model was retained as the best fit. For the analysis presented in this work, the emission-line maps shown in the figures correspond to the total component, while the presence of additional components was considered when interpreting the kinematic structure.

Following the above procedure, we produced flux, velocity, and velocity-dispersion maps for all the optical emission lines, as well as stellar kinematic maps of the continuum. For a small number of galaxies with very low-S/N emission over extended regions, a second Voronoi tessellation using VorBin \citep{Cappellari2013} was constructed using the [N II] emission-line map as the signal map . Only spatially contiguous spaxels were allowed to be combined into a bin, such that disconnected regions were not grouped together even when they had similar line fluxes. The resulting tessellation was then applied to the spectra, increasing the effective S/N while preserving the spatial structure of the ionised gas. The spectra of the resulting bins were fitted following the same procedure described above. For galaxies without detectable optical emission (see Table~\ref{tab:sample_prop}), only stellar velocity and velocity-dispersion maps were derived.

In Section \ref{sec:feedback} we use the quantities $|\sigma_{\rm los}|$ and $|v_{\rm los}|$ as global kinematic diagnostics of the warm ionized gas. According to CCA expectations, rotation-dominated systems occupy the high-$|v|$/low-$\sigma$ locus, whereas systems experiencing CCA-like turbulence and uplift populate the high-$\sigma$/low-$|v|$ regime.

Following the procedure described in \citet{gaspari18}, these values were computed by selecting all spaxels with [NII] emission detected at a significance greater than $5\sigma$ and combining them into an integrated spectrum over the full emitting region. The resulting spectrum was then fitted following the methodology described above, using between one and three Gaussian components for each emission line depending on the spectral complexity. Here $|v_{\rm los}|$ denotes the absolute shift of the integrated emission-line profile relative to the systemic velocity, while $\sigma_{\rm los}$ traces the corresponding line broadening after correction for instrumental resolution. We use these integrated measurements as ensemble-beam observables, providing a global characterization of the balance between ordered and turbulent motions for comparison with the CCA k-plot diagnostic.

\begin{table*}[t!]
\caption{\\Summary of the observations}\label{tab:observations}
\centering
\begin{tabular}{llccccccc}
\hline
\hline
LGG & BGG   & R & $z_{\text{spec}}$ & RA  & Dec  & Scale &  $\text{E(B-V)}_{\text{MW}}$  & Program ID\\
(1) &(2) & (3)& (4) & (5) & (6)  & (7)   & (8) & (9)    \\
\hline
006 & NGC 128  & 3 & 0.01411 & 00:29:15.0  & +02:51:50.8 & 0.291 & 0.0246 & 0110.A-4397(A) \\
097 & NGC 1395 & 3 & 0.00564 & 03:38:29.8  & -23:01:39.7 & 0.102 & 0.0192 & 0110.A-4397(A) \\
100 & NGC 1407 & 2 & 0.00593 & 03:40:11.9  & -18:34:49.2 & 0.112 & 0.0585 & 0110.A-4397(A) \\
126 & NGC 1779 & 3 & 0.01108 & 05:05:18.0  & -09:08:49.5 & 0.218 & 0.0757 & 0110.A-4397(A) \\
138 & NGC 2292 & 3 & 0.00750 & 06:47:39.7  & -26:44:46.5 & 0.145 & 0.1027 & 0110.A-4397(A) \\
177 & NGC 2911 & 2 & 0.01075 & 09:33:46.1  & +00:09:09.4 & 0.218 & 0.0265 & 0110.A-4397(A) \\
205 & NGC 3325 & 3 & 0.01875 & 10:39:20.5  & -00:12:01.1 & 0.388 & 0.0475 & 0110.A-4397(A) \\
236 & NGC 3665 & 2 & 0.00686 & 11:24:43.6  & +38:45:46.2 & 0.156 & 0.0164 & 0112.A-2375(A) \\
255 & NGC 3923 & 2 & 0.00604 & 11:51:02.0  & -28:48:24.7 & 0.097 & 0.0705 & 0110.A-4397(A) \\
314 & NGC 4697 & 3 & 0.004164 & 12:48:35.9  & -05:48:02.3 & 0.087 & 0.0248 & 0110.A-4397(A) \\
329 & NGC 4956 & 2 & 0.01580 & 13:05:01.0  & +35:10:41.7 & 0.344 & 0.0077 & 0112.A-2375(A) \\
341 & NGC 5061 & 3 & 0.00669 & 13:18:05.0  & -26:50:13.5 & 0.136 & 0.0591 & 0110.A-4397(A) \\
350 & NGC 5127 & 2 & 0.01625 & 13:23:45.0  & +31:33:59.8 & 0.349 & 0.0121 & 0110.A-4397(A) \\
370 & NGC 5444 & 3 & 0.01317 & 14:03:24.1  & +35:07:45.8 & 0.291 & 0.0073 & 0110.A-4397(A) \\
383 & NGC 5629 & 2 & 0.01503 & 14:28:16.4  & +25:50:55.8 & 0.325 & 0.0193 & 0110.A-4397(A) \\
398 & NGC 5903 & 3 & 0.00853 & 15:18:36.6  & -24:04:06.3 & 0.175 & 0.1295 & 0110.A-4397(A) \\
457 & NGC 7252 & 2 & 0.01588 & 22:20:44.8  & -24:40:41.9 & 0.320 & 0.0259 & 099.B-0281(A) \\
\hline
\end{tabular}{}
\tablefoot{
\noindent (1) Group name.\\
\noindent (2) BGG name.\\
\noindent (3) Richness (R) \citep{osullivan17}\\
\noindent (4) Redshift.\\
\noindent (5) Right ascension of the galaxy.\\
\noindent (6) Declination of the galaxy.\\
\noindent (7) Scale in kpc/arcseconds.\\
\noindent (8) Extinction of the Milky Way at the position of the BGG taken from dust maps of \citet{schlegel98}.\\
\noindent (9) VLT Program ID.\\
}
\end{table*}

\section{Results}\label{sec:results}

\definecolor{blush}{RGB}{252, 240, 243}
\definecolor{blush}{RGB}{254, 246, 248}
\definecolor{blush}{RGB}{250, 235, 240}

\begin{figure*}[t]
\centering

% Panel 1: Filaments
\begin{tcolorbox}[
    colback=white,
    colframe=white,
    boxrule=0.8pt,
    arc=2pt,
    boxsep=3pt,
    title={Filament systems},
    colbacktitle=gray!10,
    coltitle=black,
    fonttitle=\bfseries
]
\noindent
\includegraphics[width=0.49\linewidth]{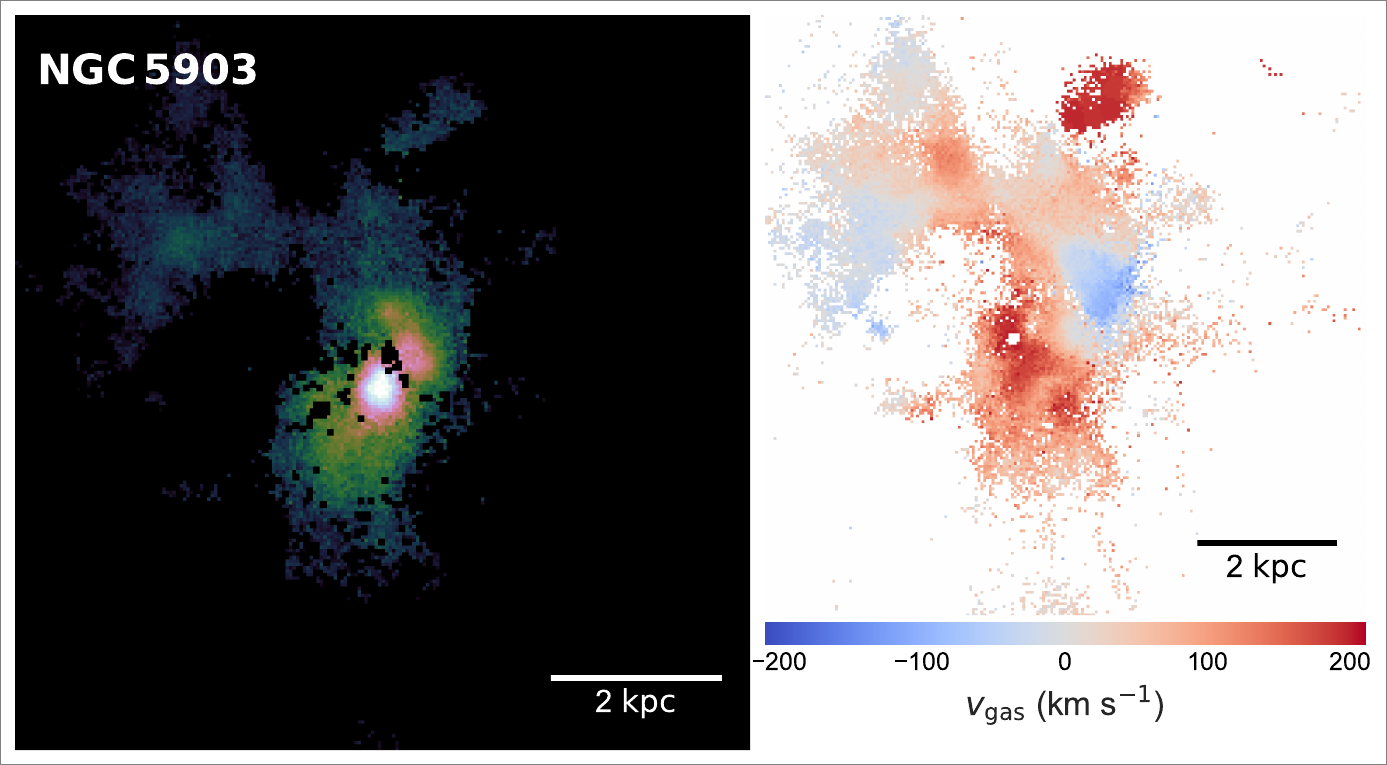}%
\hfill
\includegraphics[width=0.49\linewidth]{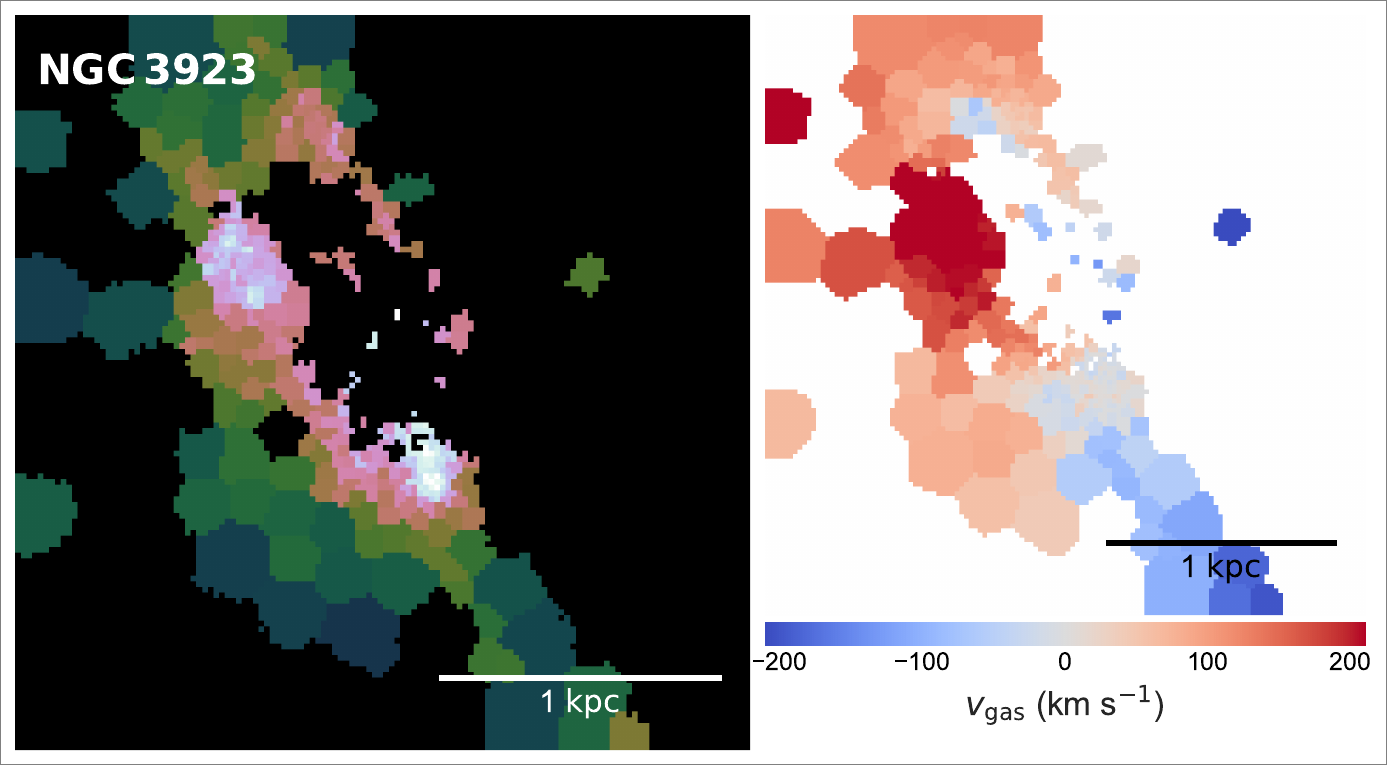}
\end{tcolorbox}

% Panel 2: Disks
\begin{tcolorbox}[
    colback=white,
    colframe=white,
    boxrule=0.8pt,
    arc=2pt,
    boxsep=3pt,
    title={Extended and compact disks},
    colbacktitle=gray!10,
    coltitle=black,
    fonttitle=\bfseries
]
\noindent
\includegraphics[width=0.49\linewidth]{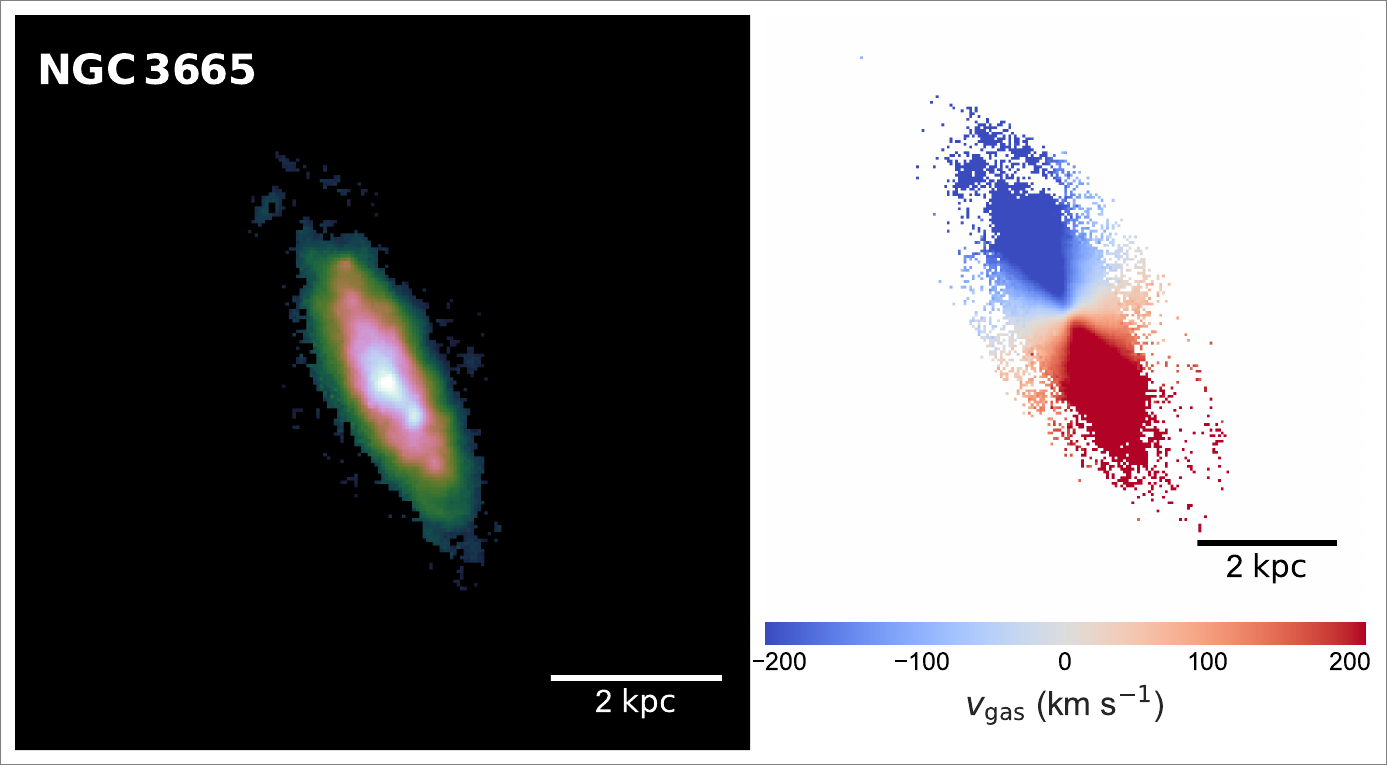}%
\hfill
\includegraphics[width=0.49\linewidth]{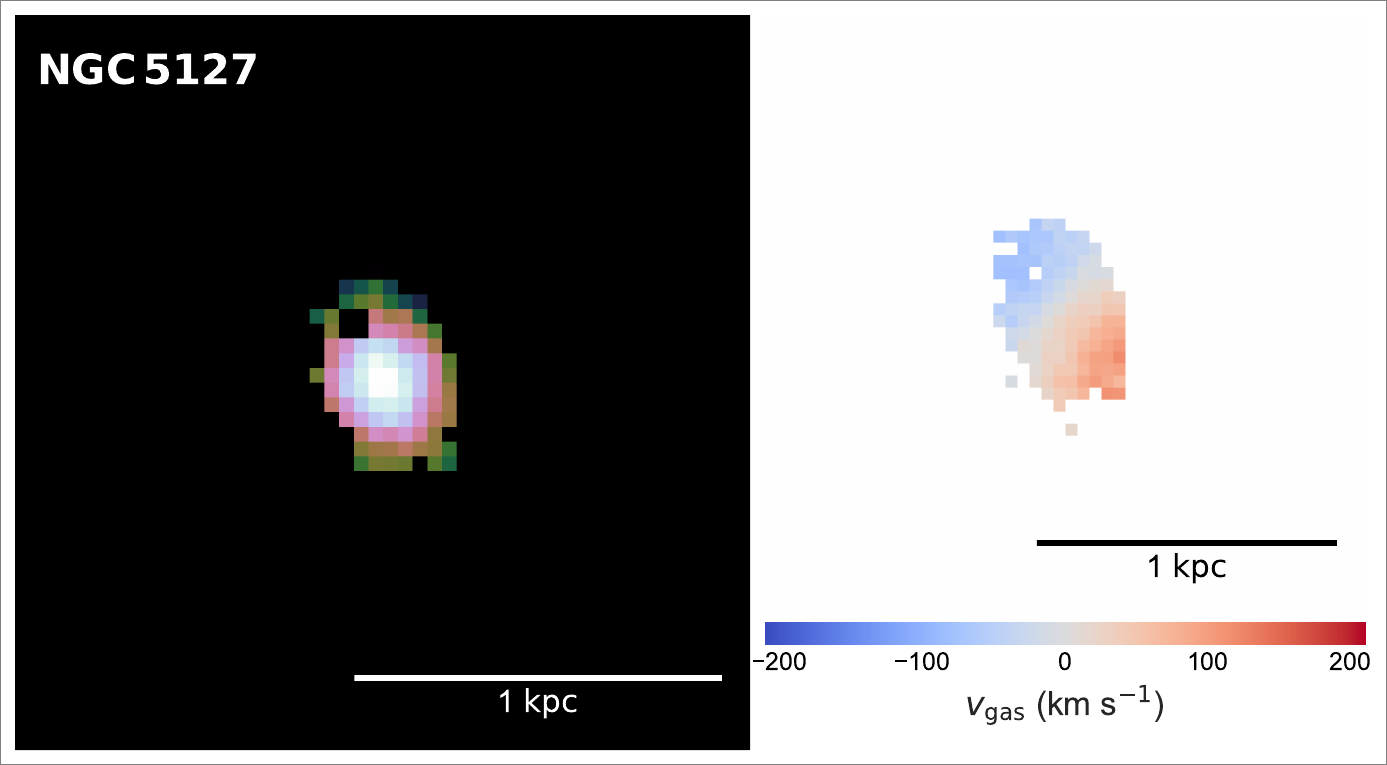}
\end{tcolorbox}

% Panel 3: Rings
\begin{tcolorbox}[
    colback=white,
    colframe=white,
    boxrule=0.8pt,
    arc=2pt,
    boxsep=3pt,
    title={Rings},
    colbacktitle=gray!10,
    coltitle=black,
    fonttitle=\bfseries
]
\centering
\includegraphics[width=0.49\linewidth]{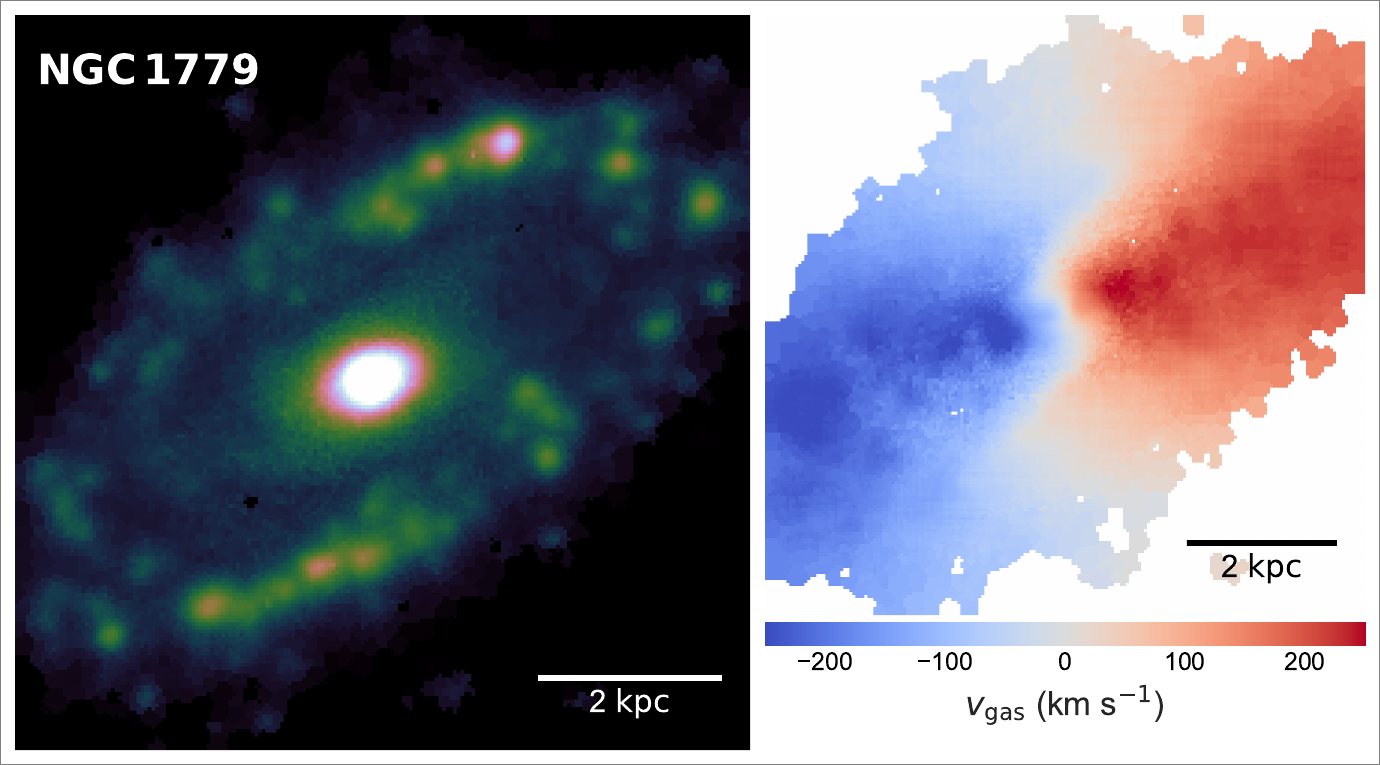}
\end{tcolorbox}

\caption{Example of the diverse morphology (first column) and kinematics (second column) of the warm ionized gas of our MUSE sample. First row: example of sources with asymmetric gas distribution, and disturbed velocity maps, categorized as Filaments. Second row: example of sources displaying regular rotation in the form of extended (rotation-dominated) and compact (dispersion-dominated) disks. Last row: example of sources displaying rings.}
\label{fig:gas_distribution}
\end{figure*}

\subsection{Warm ionized gas}
The MUSE observations of our sample reveal that 13 of 17 galaxies (77\%) have optical emission lines ([NII] or H$\alpha$). The warm ionized gas traced by optical emission lines reveals a diverse set of morphologies and distributions, traced primarily by H$\alpha$ emission.
It is worth noticing that [NII]$\lambda$6583 was used to identify the presence of warm ionized gas when H$\alpha$ was too faint compared to the stellar continuum. A detailed description of the distribution and kinematics of each source is provided in Appendix \ref{sec:fullsample}.

We classify the warm ionized gas morphologies and kinematics into three categories based on a combination of spatial structure and velocity coherence: disks, rings, and filamentary/complex systems. For that aim, we characterize and separate the kinematic regimes of the sources, and compute a flux-weighted global $V/\sigma$ for each system, as follows,
\begin{equation}
    \frac{V}{\sigma} = \sqrt{\frac{\sum F_{i}V_{i}^{2}}{\sum F_{i}\sigma_{i}^{2}}},
\end{equation}
where $i$ corresponds to each spaxel of the velocity, $V$, velocity dispersion, $\sigma$, and H$\alpha$ or [NII] flux maps, respectively. Galaxies with $V/\sigma > 1$ are rotation-dominated, whereas those with $V/\sigma < 1$ are dispersion-dominated. We emphasize that $V/\sigma$ is used as a global kinematic indicator rather than a strict morphological classifier. In particular, systems with clear disk-like morphologies but low global $V/\sigma$ values may reflect beam smearing, low inclination, or dynamically hot disks. In Figure \ref{fig:gas_distribution} we present examples of such categories: filamensts/complex systems (left panel), extended disks/rings (rotation dominated, right upper panel) and compact disks (dispersion dominated, right bottom panel).

In Figs. \ref{fig:Halpha_distribution}-\ref{fig:Halpha_distribution7}, in Appendix \ref{sec:fullsample}, we display and describe the spatial distribution and kinematics of the warm ionized gas and the stellar component for the entire sample. In Table \ref{tab:sample_prop} we describe the distribution of the warm ionized gas and the size for each source. Below we describe each category in more detail:\\

\noindent\textbf{Extended disks and rings (rotation-dominated)}
Rotation-dominated disks are characterized by disk or ring-like morphologies, a coherent and monotonic large-scale velocity field, and global kinematics dominated by ordered rotation ($V/\sigma > 1$). In these systems, the emission-line maps show a smooth, centrally concentrated morphology that is roughly symmetric with respect to the galaxy center, but some sources also show tails or diffuse gas. The velocity fields exhibit a continuous, large-scale gradient. The velocity dispersion is typically lower along the disk plane and peaks toward the center, as expected for rotationally supported systems. 
Based on these characteristics, the galaxies hosting clear rotating disks or rings are NGC 1779, NGC 2292, NGC 2911, NGC 3665, NGC 4697, NGC 4956, and NGC 5444.

In particular, NGC 5444, and potentially NGC 1779, we noted a modest central kinematic twist of $\sim45^\circ$ produces an apparent deviation from a simple rotation pattern, such twists are also observed in otherwise regular disk galaxies and can arise from internal secular processes, such as stellar bars or secular distortions.\\

\noindent\textbf{Compact disks (dispersion-dominated)}
Dispersion-dominated disks also exhibit disk-like morphologies and coherent velocity gradients, but with higher velocity dispersions such that random motions contribute significantly to global kinematics ($V/\sigma < 1$). In these systems, the low $V/\sigma$ likely reflects a combination of factors such as disk thickness, turbulence, or low inclination rather than the absence of ordered rotation. Two systems show such characteristics, NGC 5061 and NGC 5127. These two sources are also the only compact disks of the sample, with sizes 0.2 -- 1.6 kpc.

We note that the spatially resolved BPT diagram of NGC 5061 indicates LINER-like emission, abd we find no clear evidence for AGN ionization (Olivares et al in prep).\\

\noindent\textbf{Filaments and complex systems}. Such systems show irregular or asymmetric morphology, often tracing radial, but also curved structures and tails. Their velocity fields lack coherent velocity gradients along the entire warm ionized gas distribution, and instead show a complex or an asymmetric velocity map.  
%We also classified as filamentary sources that show a sort of symmetric velocity distribution but also display extended structures (NGC128, NGC7252).
Extended rotating structures are distinguished from filaments with projected symmetry based on the coherence and spatial continuity of the velocity field. Rotating structures exhibit a large-scale, monotonic velocity gradient across the full extent of the structure, a well-defined kinematic center. In contrast, filaments with projected symmetry show localized or non-monotonic velocity gradients, lack a clear kinematic center, consistent with disturbed or non-axisymmetric kinematics rather than ordered rotation.
Examples of filamentary systems include NGC 128, NGC 3923, NGC 5903, and NGC 7252.\\

In particular, NGC 7252, provides an interesting example in which the ionized gas appears to comprise two kinematically distinct components. A central region hosts a well-defined inner disk, characterized by a regular velocity gradient. The inner disk, is surrounded by extended filamentary structures showing a complex morphology and kinematic. Given the disturbed merger history of NGC 7252, the filament may represent residual tidal material or gas associated with the ongoing settling of the system, while the central disk has already reached a more dynamically ordered state.

\noindent\textbf{Non-detections}:
A few galaxies, such as NGC 1395 and NGC 1407, show no detectable warm ionized gas despite hosting extended X-ray halos. This may indicate that AGN feedback has heated or expelled the cooling gas, preventing the formation of filaments and complex structures or disks (see more details and discussion below in Section \ref{sec:discussion}).

Based on this classification, we found that at least four sources display filamentary and complex structures similar to those seen in the centers of clusters but on smaller scales. In contrast, most systems with warm ionized gas (9 of 13) reveal disks.

\subsection{Stellar Component}
The stellar continuum maps display a wide diversity of morphologies and structural features across our sample. 
We classify the stellar kinematics as either regularly rotating (RR) or non-regularly rotating (NRR) based on visual inspection of the velocity fields (see Table \ref{tab:pa_fit}). The majority of galaxies with ionized gas (11 of 13) exhibit clear large-scale stellar rotation, and thus are classified as RR. Some of these RR galaxies, such as NGC 7252 and NGC 5061, display evidence of double stellar disks, suggesting complex dynamical structures possibly linked to past merger event (see Section \ref{sec:discussion}).

Only NGC 5903 and NGC 3923 display disturbed stellar velocity fields lacking clear large-scale rotation, and were classified as NRR, similar to the kinematic patterns reported for central galaxies in massive groups and clusters \citep{Loubser2022,hamer14}. Even in these cases, subtle hints of ordered motion are observed in the central regions.

\begin{table*}[t!]
\caption{\\Sample properties.}\label{tab:sample_prop}
\centering
\footnotesize
\makegapedcells
\begin{NiceTabular}{lcccccccccc}
\hline
\hline
BGG & type & X-ray & Radio  & $\rm M(HI)$& $\rm M(H2)$ & $\rm log\, (M_\star)$ & min($t_{\rm cool}/t_{\rm ff}$)& H$\alpha$  & Size & L(H$\alpha$)\\ 
  &  &  & & (10$^{8}$\,$\rm M_{\odot}$) & (10$^{8}$\,$\rm M_{\odot}$) & ($\rm M_{\odot}$) &  & Morphology &      (kpc)  & (10$^{39}$ erg s$^{-1}$) \\ 
 (1)  &  (2) & (3) & (4) & (5) & (6) & (7) &    (8) &         (9) & (10) & (11) \\ \hline
NGC 128 & S0 & GRP &    point   & 4.46 &  1.76$\pm$1.5& 11.09 & 54.1$\pm$23.0 & filaments &  16   & 13.1 $\pm$ 1.5  \\
NGC 1395 & E & GRP & point & $<$2.81 &$<$0.27 & 10.91 &  50.2$\pm$1.9 & NE &    --   & -- \\
NGC 1407 & E &  GRP & remnant & $<$3.71 & $<$0.34 & 11.09 & 22.7$\pm$9.5 & NE & --  & --\\
NGC 1779 & S0-a & gal & point & 18.19 & 4.57$\pm$0.60 & 10.75 & 33.3$^{+7.3}_{-5.0}$ & inner disk + outer ring &    11.5   & 2752.8 $\pm$ 153.4  \\
NGC 2292 & S0& -- & --    & 22.38 & $<$0.47 & 10.79 & -- & ring &     1.70   & 1.1 $\pm$ 0.2 \\
NGC 2911 & S0 & GRP & point & 20.41 & 2.66$\pm$0.31 & 10.82 & 105.8$\pm$8.0 & disk + tails    &       10.5 & 13.2 $\pm$ 0.4  \\
NGC 3325 & E & -- & -- & 15.13 & $<$1.57 & 10.88 & -- & NE      &    --  & --   \\
NGC 3665 &S0 & gal & small-scale jet & $<$0.21 & 6.0$\pm$0.32 & 10.72 & 43.6$\pm$2.7 & disk     &    5.5     & 738.5 $\pm$ 9.6\\
NGC 3923 & E & gal  & point & $<$2.57  & $<$0.29 &    11.03 & 25.3$\pm$0.6& filaments     &      3  & 0.013 $\pm$ 0.004 \\
NGC 4697 & E & GRP & point & $<$1.38 & $<$0.07 & 10.93 & 58.3$\pm$2.5 &   disk + diffuse gas     &   3.5     & 25.2 $\pm$ 0.3 \\
NGC 4956 & S0&  gal & -- & 11.22 & $<$0.75 & 10.91 & 98.7$^{+40.7}_{-51.6}$ & disk + tails& 10 & 6.0 $\pm$ 0.8\\
NGC 5061 & E& -- & point & $<$4.57 &$<$0.43 & 10.96 & -- & compact disk & 0.42 & 12.2$\pm$ 10.0\\
NGC 5127 & E & GRP & large-scale jet  & 47.86 &  0.77$\pm$0.05& 10.96 & 105.1$\pm$23.8& compact disk &    1.5   & 2.4 $\pm$ 0.2  \\
NGC 5444 & E & GRP & -- & $<$6.16 & $<$0.64 & 11.04 & 59.3$\pm$34.7 & disk + diffuse gas & 7.5 & 5.3 $\pm$ 0.1 \\
NGC 5629 & S0 & GRP & -- & -- & $<$0.91 & 10.95 &24.1$\pm$5.1 & NE & -- & --\\
NGC 5903 & E & GRP & diffuse & 30.19 & $<$0.58           &  10.73 & 67.1$\pm$4.2 & filaments & 10.5 & 99.9 $\pm$ 4.7\\
NGC 7252 & S0 & gal & point &  40.73 & 58.0$\pm$8.70 & 10.92 & 10.7$\pm$12.1 & fils/tidal/inner ring & 36 & 279.1 $\pm$ 0.1\\ 
\hline
\end{NiceTabular}
\tablefoot{
\noindent (1) BGG name.\\
\noindent (2) Galaxy morphology type from HyperLeda.\\
\noindent (3) X-ray morphology. GRP: Group, gal: galaxy (O'Sullivan et al. in prep).\\
\noindent (4) Radio distribution \citep{Kolokythas2019}.\\
\noindent (5) HI masses are drawn from \citep{osullivan18b}.\\
\noindent (6) Cold molecular gas masses are drawn from \citet{osullivan18b} and  from \citet{osullivan15}.\\
\noindent (7) Stellar mass \citet{osullivan18b} and  from \citet{osullivan15}.\\
\noindent (8) minimum ratio of isochoric cooling time versus free-fall time (O'Sullivan et al. in prep).\\
\noindent (9) Distribution of the optical emission lines. NE: No emission.\\
\noindent (10) Size of the optical emission lines (H$\alpha$, [\textsc{NII}]$\lambda6584$ emission).\\
\noindent (11) H$\alpha$ luminosity.
}
\end{table*}

\section{Discussion}\label{sec:discussion}
In this section, we discuss the results obtained from our new MUSE observations. One of the main goals of this paper is to explore how these group-dominant galaxies acquired the cold and warm ionized gas. We discuss the different scenarios: stellar-mass loss, mergers and interactions, and cooling from their hot halos.

\subsection{Comparison of the gas and stellar kinematics}\label{sec:PA}
\begin{figure}[ht!]
    \centering
    \includegraphics[scale=0.8]{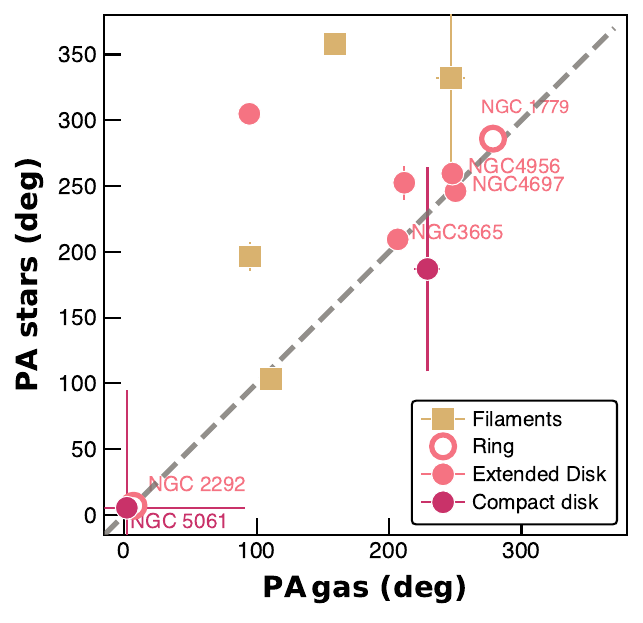}
    \caption{Comparison of the position angle of the stars and the gas. Symbols indicate galaxies with distinct gas morphologies: extended disks (pink circles), compact disks (dark pink), rings (open pink circles), and filaments (yellow squares). The dashed line marks the one-to-one relation.}
    \label{fig:PA}
\end{figure}

For the sources with detected ionized gas, we compare the kinematics of the warm ionized component with those of the stellar population. To enable a more quantitative comparison, we measure the position angle ($PA$) from the velocity maps using the {\sc $fit\_kinematic\_pa$} routine developed by \citet{2005MNRAS.357.1113K}. The PA is defined as the angle measured counterclockwise from the projected north direction to the location of the maximum positive velocity. All the values are listed in Table \ref{tab:pa_fit}).

Following \citet{Olivares2022}, we define the degree of misalignment taking the difference between both $PA$, stellar and gaseous, $\rm \Delta PA = PA_{\star} - PA_{gas}$, and define a target as  misaligned if this difference is greater than $\rm \Delta PA>30\deg$. The large uncertainties in the gas position angle for some sources classified as filaments and complex systems arise from their disturbed velocity fields. For filamentary systems, the concept of a single kinematic position angle is less physically meaningful, and large $\Delta PA$ values primarily reflect disturbed, non-axisymmetric gas motions. Also due to the asymmetry of the velocity field of filamentary system the $PA$ may be uncertain. Therefore $\Delta PA$ is more significant for coherent rotating systems, while for filaments and complex systems, the decoupled classification reflects disturbed kinematics more than a strict $PA$ metric.

In Figure \ref{fig:PA}, we present the position angles of the stellar and gaseous components for our sample. The dashed line corresponds to the one-to-one relation. As depicted in Fig. \ref{fig:PA}, 6 galaxies exhibit gaseous disks that are kinematically aligned with the stars, namely NGC 1779, NGC 2292, NGC 3665, NGC 4697, and NGC 4956, NGC 5061, with $\Delta \mathrm{PA}<20^\circ$, consistent with the threshold adopted in previous studies \citep{Olivares2022,2015MNRAS.448.1271L}.

Additionally, in most of the filamentary sources (yellow rectangles) the gas and the stars are decoupled. The exception being NGC 7252, which has a $\Delta \mathrm{PA} < 20^\circ$, but the overall kinematics of both components are not coupled, as indicated by their distinct velocity fields (see Fig. \ref{fig:Halpha_distribution}). 

\begin{table}[htb]
\caption{Position angle of the ionized gas rotating disks and stellar disks.}
\centering
\begin{tabular}{lcccc}
    \noalign{\smallskip} \hline \hline \noalign{\smallskip}
   \footnotesize BGG & rotation &  PA$_{\rm gas}$ & PA$_{\rm stellar}$ & $\rm \Delta PA$\\
   \footnotesize  & RR/NRR &  ($\degr$) & ($\degr$) & ($\degr$)\\
    \noalign{\smallskip} \hline \noalign{\smallskip}
    {\it Filaments }   &  &  &  & \\
    NGC 128 & RR &159.5$\ \pm\ $0.5  &358.0$\ \pm\ $1.5  & 204.0 \\
    NGC 3923 & NRR &246.5$\ \pm\ $10.8  &332.0$\ \pm\ $89.8  & 85.5\\
    NGC 5903 & NRR &95.0$\ \pm\ $0.5  &196.5$\ \pm\ $10.8 & 101.5 \\ 
    NGC 7252 & RR &111.0$\ \pm\ $0.5  &103.5$\ \pm\ $0.5 & 7.5 \\
    {\it Disks}   &  &  &  & \\
    NGC 1779 & RR &278.5$\ \pm\ $0.5  &286.0$\ \pm\ $0.5  &7.5 \\
    NGC 2911 & RR &94.5$\ \pm\ $0.5  &304.9$\ \pm\ $0.5  & 210.4 \\
    NGC 2292 &RR  &7.0$\ \pm\ $1.2  &7.0$\ \pm\ $1.8  &0.0 \\
    NGC 3665 & RR &206.5$\ \pm\ $0.5  &209.5$\ \pm\ $0.5  &3.0 \\ 
    NGC 4697 &RR &250.2$\ \pm\ $2.0  &246.1$\ \pm\ $1.8  & 6.0 \\
    NGC 4956 &RR &247.8$\ \pm\ $0.5  &259.5$\ \pm\ $3.0 & 12.5 \\
    NGC 5061 & RR &2.0$\ \pm\ $89.0  &5.5$\ \pm\ $89.5  &3.0 \\ 
    NGC 5127 & RR &229$\ \pm\ $9.8  & 187.0$\ \pm\ $77.8  & 42.0 \\ 
    NGC 5444 & RR &211.5$\ \pm\ $3.8  &252.5$\ \pm\ $13.0  &41.0 \\
    \noalign{\smallskip} \hline \noalign{\smallskip}
\end{tabular}
\tablefoot{
\noindent Rotation of the stellar component (RR: Regular Rotation, NRR: Non Regular Rotation).
\raggedright The kinetic PA is measured anticlockwise from the north direction to the maximum positive velocity. The errors on the kinetic PA are 3$\sigma$ errors.\\
\raggedright $\rm \Delta PA$: Difference between the stellar and gas kinematic position angles, $\rm \Delta PA = | PA_{\rm stellar} - PA_{\rm gas} |$\\
}
\label{tab:pa_fit}
\end{table}

\subsection{Origin of the warm ionized gas}

In this section, we discuss how the group-dominant galaxies acquire their cold gas based on their warm gas and stellar kinematics and distribution, and the properties of their X-ray halos and jets. In Table \ref{tab:summary_origin} we summarize our findings.  
Our interpretation is that the origin is guided by a decision tree in which (1) kinematics and distribution of the gas provides the primary discriminants, (2) followed by the coupling/decoupling of the gas and stars, (3) presence of merger signatures, (4) the thermodynamic state of the hot halo, (5) evidence for radio–jets, (5) finally we used the relative cold gas mass budget ($M_{\rm stellar\_mass\_loss}$ versus $M_{\rm gas}$) as supporting diagnostics.

\subsubsection{Stellar mass loss}\label{sec:stellarmassloss}

\begin{figure}[ht!]
    \centering
    \includegraphics[scale=0.8]{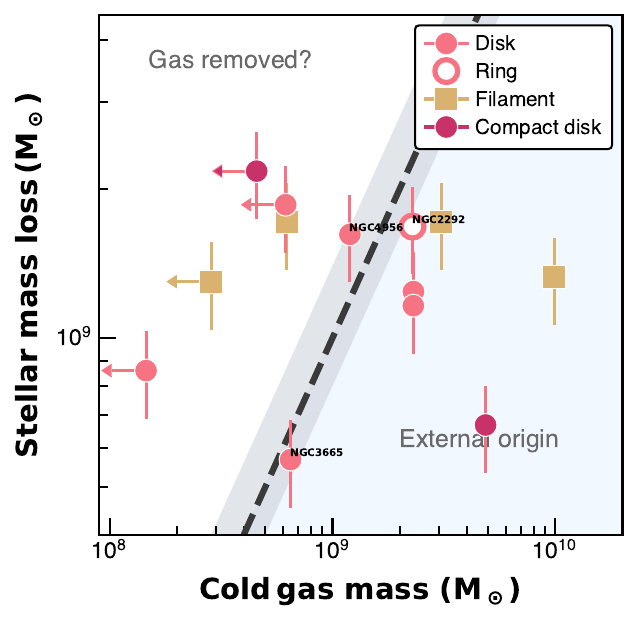}
    \caption{Relationship between the observed cold plus warm gas mass and the predicted gas mass from stellar mass loss for early-type galaxies in our sample. Symbols indicate galaxies with distinct gas morphologies: disks (pink circles), disks/rings (open pink circles), darker pink (compact disks), and filaments (yellow squares). The right arrows correspond to sources with upper limits on both HI and H$_2$ detections. The dashed line marks the one-to-one relation, while the shaded region indicates the 0.25 sigma range around it. The light blue area highlights the regime where the cold gas is expected to be of external origin (e.g., acquired through accretion or minor mergers, or cooling from the hot halos), whereas the white region corresponds to internal recycling or galaxies having experienced substantial gas removal. Galaxies lying close to the one-to-one relation are consistent with an origin dominated by stellar mass loss. Representative galaxies lying in these regions are labeled for clarity.}
    \label{fig:stellar_mass_loss}
\end{figure}

Stellar mass loss from evolved stars provides a source of gas in early-type galaxies. As asymptotic giant branch (AGB) stars and red giants evolve, they shed material into the ISM, enriching it with gas and metals. In the absence of substantial ongoing star formation, this process can accumulate enough material to form the observed warm ionized gas. Stellar mass loss injects a metal-rich gas, but this gas could be rapidly shock-heated, mixed with the ambient hot medium, and thermalized to the local virial temperature of the surrounding gas \citep{Parriott2008}. 

Eventually, under the right thermodynamical conditions, this hotter gas can cool and condense (see more in Sect. \ref{sec:cooling}), in time-scale of $\sim 10^6$ yr \citep{Mathews1990}. It is worth mentioning that once the gas gets thermalized, it becomes observationally indistinguishable from the surrounding hot halo, and any gas that subsequently cools cannot be unambiguously associated with either stellar mass loss or the pre-existing hot atmosphere. Although a fraction of  $\sim$25\% may remain cold from the stellar ejecta, according to calculations by \citet{Parriott&Bregman2008}. More updated studies, \citet{Bregman2009} argue that if radiative cooling is included, then most of the material ($>$80\%) remains cool, does not mixed with the hotter gas.

The properties of the cold and warm gas that is ejected through a stellar mass loss, such as its spatial distribution, kinematics, and metallicity, are expected to reflect the stellar population from which it originates. For example, gas produced via stellar mass loss should follow the stellar light, exhibit low relative velocities with respect to the stars, and similar metallicities. In the same way, aligned rotating disks with a smooth morphology are consistent with internal recycling, while strong misalignment and filamentary morphologies typically require external supply and/or condensation.  

One way to test the stellar mass loss scenario as a source of the warm ionized gas in our sample is to examine whether the gas kinematics and spatial distribution are coupled to those of the stars. Most galaxies in our sample exhibit stellar disks, so if stellar mass loss dominates, we would expect the gas to form co-rotating, disk-like structures aligned with the stellar component.

In Section \ref{sec:PA}, we compared the kinematic position angles of the gas and stellar components. Our analysis reveals that six galaxies: NGC 1779 (ring), NGC 2292 (ring), NGC 3665, NGC 4697, NGC 4956, and NGC 5061, host gaseous disks that are kinematically coupled with their stellar counterparts. This kinematic alignment suggests that the warm ionized gas is dynamically coupled to the stellar component, favoring an internal origin. However, given that our analysis probes only the central $\sim$5–10 kpc, where dynamical timescales are short, externally accreted gas may also have had sufficient time to settle into a relaxed disk. It is worth noting that some of these disks exhibit additional structures, including an outer ring in NGC 1779, diffuse ionized gas in NGC 4697, and extended tail-like structures in NGC 4956. These features are not necessarily kinematically decoupled from the central gas disks; rather, their presence suggests that an origin solely from ongoing stellar mass loss may not be unique. While stellar mass loss appears sufficient to explain some of aligned disks, it fails to account for the filamentary systems or misaligned disks, which require external or cooling-driven accretion. This mixed picture is also consistent with broader studies across galaxies, groups, and clusters showing that, particularly in lower-cooling-rate systems, recycled stellar ejecta can provide a non-negligible contribution to the cool gas reservoir alongside hot halo cooling (e.g., \citealt{McDonald2018}).

Estimating the mass of the colder gas and comparing it with predictions from integrated stellar mass loss rates offers valuable insight into the mechanisms through which the galaxies in our sample acquired their gas. To explore this, we needs the total cold and warm gas budget ($M_\mathrm{gas}$). First, we estimate the mass of warm ionized gas in our sample from the observed H$\alpha$ luminosity ($L_\mathrm{H\alpha}$, see Table \ref{tab:sample_prop}), adopting standard recombination relations and assuming typical electron densities ($n_{\rm e} \sim 100~\mathrm{cm}^{-3}$, e.g. \citealt{osterbrock89}). To account for the total content of the cold gas ($M_{\rm gas}$), we also include estimates of the masses of the cold and neutral molecular gas of \citet{osullivan18b}, since the majority of the gas is expected to reside in these phases. 

The resulting gas masses ($M_{\rm gas}$) can then be compared to the total mass expected from stellar mass loss ($ M_\mathrm{stellar\_mass\_loss}$) over a few gigayears. 
We estimated the accumulated mass as follows:
\begin{equation}
    M_\mathrm{stellar\_mass\_loss} = \dot{M_{\star}} \times \Delta t
\end{equation}
where $\dot{M_{\star}} \sim 1.5 \times 10^{-11} (L_\mathrm{B}/L_\odot)~M_\odot \mathrm{yr^{-1}}$ (\citep{Ciotti1991}, see also \citealt{LeitnerKravtsov2011}, and $L_\mathrm{B}$ is the galaxy B-band luminosity. The $L_\mathrm{B}$ luminosity was taken from the HyperLEDA database \citep{makarov14}. We adopt a timescale of $\sim$2.5 Gyr, with an uncertainty of 0.5 Gyr, comparable to the cooling time of the hot halo at intermediate radii, where gas can accumulate before becoming thermally unstable \citep{osullivan17}. Shorter timescales (e.g. 1 Gyr) would reduce the accumulated mass, but would not affect our conclusions. 

If $M_\mathrm{gas}$ is comparable to $M_\mathrm{stellar\_mass\_loss}$, this supports an internal origin of the warm and cold gas via stellar ejecta, while significant deviations may indicate additional processes such as cooling from hot halos, external accretion from mergers and interactions, or gas removal by AGN feedback.

Systems located above the one-to-one line indicate cases where internal processes (Fig.\ref{fig:stellar_mass_loss}), such as stellar or AGN feedback have efficiently removed part of the gas. On the other hand, systems located below the one-to-one relation, suggest that a significant fraction of their cold gas reservoirs are of external origin, possibly supplied by minor mergers or accretion events. We note that while this diagnostic is useful for distinguishing broad regimes, it is subject to substantial systematic uncertainties (e.g., gas-phase partitioning, and the use of upper limits), and should therefore be interpreted as an order-of-magnitude discriminator rather than a precise classifier, particularly given our small sample size. As a result, Fig. \ref{fig:stellar_mass_loss} is best viewed as a consistency check rather than a decisive discriminator between gas origin scenarios.

Three disk systems are located near the one-to-one relation (NGC 2292, NGC 3665, and NGC 4956). These systems also have stellar and gas coupled kinematics ($\Delta \mathrm{PA} \approx 0\degr$) that support an internal origin for the interpretation.

Of the systems with coupled gas and stars, we have NGC 1779, disk with an outer ring source ($\Delta \mathrm{PA} \approx 0\degr$), shows a $M_\mathrm{gas}>>M_\mathrm{stellar\_mass\_loss}$ suggesting some contribution from an external source. Such an excess is consistent with gas accreted in a past interaction that has since re-aligned with the stellar disk over 1–2 Gyr (see following section), thereby preserving the observed kinematic coupling while supporting an external origin. NGC 4697 and NGC 5061 only upper limits are available for their cold gas mass, making the interpretation more uncertain. 

Two filamentary systems and ring sources, have $M_\mathrm{gas}/M_\mathrm{stellar\_mass\_loss}>>1$, suggesting that additional sources of gas, such as external accretion or cooling, are likely responsible for the formation of the warm gas.

Lastly caution must be taken to interpreter stellar-mass loss as the origin for systems with $\Delta \mathrm{PA} \approx 0^\circ$, as these do not necessarily indicate an internal origin, but may instead reflect past accretion that has since settled into co-rotation as simulation shown externally accreted gas can be initially misaligned with the stellar component, but re-aligns on $\sim 1$–$2,$Gyr timescales \citep[e.g.,][]{Davis2011,vandeVoort2015}.

\subsubsection{Mergers, and satellite interactions}

\begin{figure*}[t]
    \sidecaption
    \includegraphics[width=5.9cm]{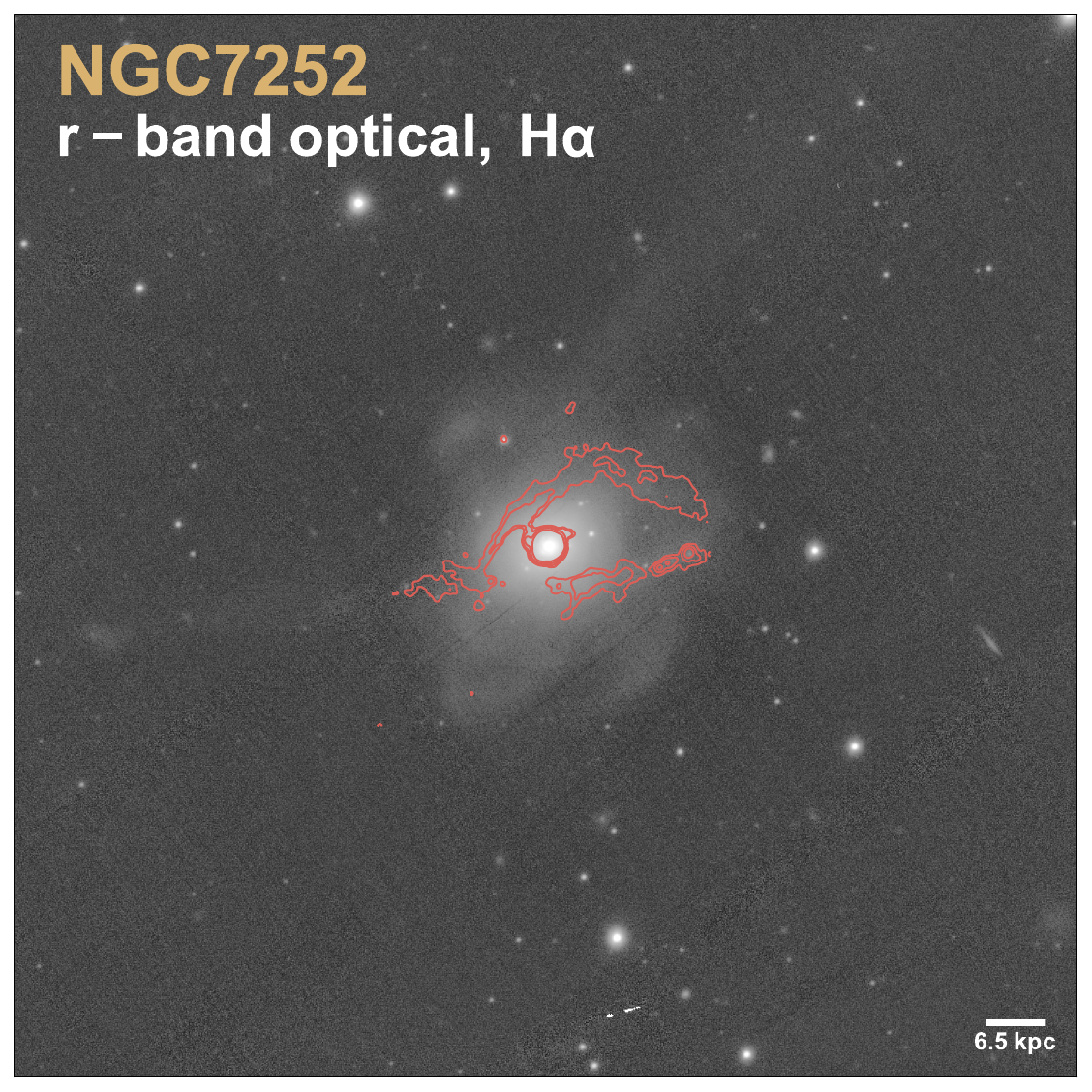}
    \hspace{0.2cm}
    \includegraphics[width=5.9cm]{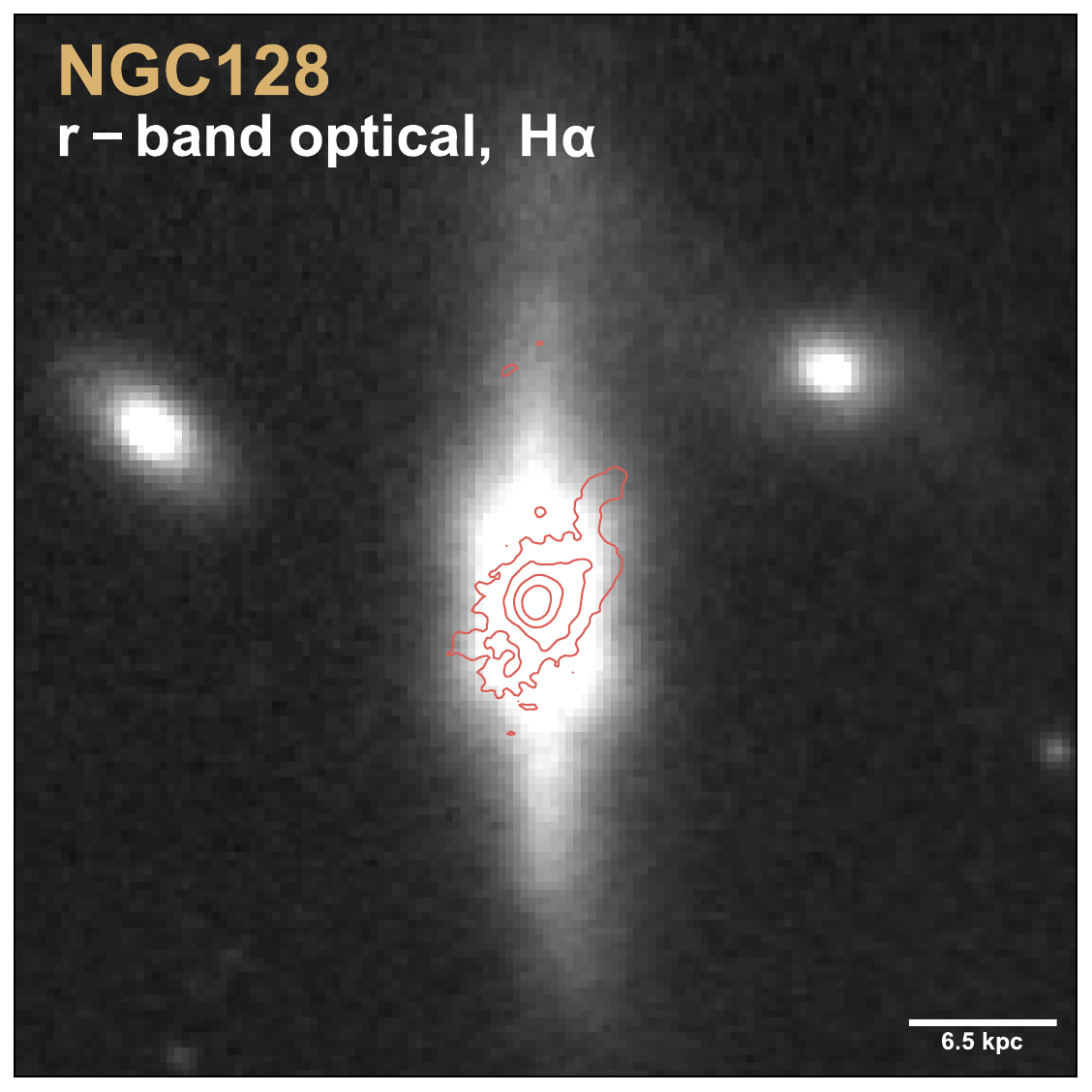}
    \caption{Optical images of galaxies in the sample that have
    undergone a merger or are interacting, with H$\alpha$ contours
    shown in red. From left to right: NGC~7252 and NGC~128.}
    \label{fig:interacting}
\end{figure*}

After galaxy mergers and interactions, the formation of a gaseous disk depends primarily on the gas content of the progenitor galaxies, the merger mass ratio, and the dynamics of the interaction \citep{Barnes1991,Lotz2010}. In major mergers, pre-existing disks are often disrupted, and gas loses angular momentum, funneling toward the central regions where it can fuel starbursts or AGN activity \citep{Hopkins2008}. However, residual gas that retains sufficient angular momentum after the merger, as well as gas re-accreted from tidal tails or the surrounding intergalactic medium, can settle into a rotating disks over timescales of 1–2 Gyr \citep{Hopkins2009}. On the other hand, minor mergers are generally less disruptive, allowing gas from the secondary galaxy to be settled into the disk. Observationally, post-merger galaxies often exhibit extended gaseous disks, sometimes misaligned with the stellar component, suggesting either residual or externally acquired gas. 

In our sample, three disks systems exhibit decoupled stellar and gas kinematics (NGC 2911, NGC 5127 (compact disk), and NGC 5444). Our analysis presented in Section~\ref{sec:stellarmassloss} further supports an external or partially external origin for NGC 5127 and NGC 2911, likely through satellite interaction or cooling from the hot halo, as the mass expected from stellar mass loss is insufficient to account for the observed cold gas reservoir. In particular NGC 2911 (disk) has counter-rotation between the gas and the stars, indicating an acquisition for the gas through a minor merger or accretion event, rather than internal stellar mass loss.

As mentioned, some gaseous disks can settled in the gravitational potential of the stellar component after a couple of Gyr. Indications of past interaction or mergers would be imprinted in the distribution of the gas in the form of streams and tails. In our sample, such features are observed in NGC 4697 and NGC 4956, as well as in the kinematically decoupled systems NGC 2911 and NGC 5444.

Another signature of past merger activity or accretion event can be inferred from the stellar kinematics, particularly through the detection of a kinematically decoupled core (KDC) \citep{2008MNRAS.390...93K,Emsellem2007,2014MNRAS.444.3357N,1992A&A...258..250B}. Within our sample, we identify such structures in three galaxies: NGC 1779 (disk/ring); NGC 5061 (compact disk); and NGC 7252 (filaments, tidal tails, and nuclear ring). While, the KDC records a merger in the past, it does not necessarily mean that the present-day ionized or molecular gas still traces that same event. 

Ring-like structures around early-type galaxies are often indicative of external gas accretion \citep{Finkelman2011}, potentially arising from the capture of gas during minor mergers or satellite interactions, which can imprint distinct kinematic signatures. In particular, NGC 1779 exhibits an inner gaseous disk and an outer clumpy ring, while NGC 2292 displays a ring. The presence of both an inner disk and an outer ring suggests a multi-phase accretion history, where the inner component may have settled into rotation, while the outer ring traces more recent or ongoing infall.

One interesting example, is NGC 7252 (filaments, tidal tails, and inner ring) which shows clear evidence of previous galaxy merger events as displayed by optical images (see Fig. \ref{fig:interacting}). NGC 7252, also known as the ''Atoms-for-Peace galaxy´´, is the BGG in the X-ray faint LGG 457 galaxy group. It is the remnant of a major merger between two gas-rich spirals, exhibiting two prominent HI-rich tidal tails \citep[e.g.,][]{1994AJ....107...67H}. In addition, the H$\alpha$ filaments follow those tidal tails.  

There are also a few systems that show signs of closed interaction: NGC 128 is in tidal interaction with NGC 127, and potentially the nearby NGC 130. One of the H$\alpha$ tails of NGC 128 extends in the direction of the companion galaxy, suggesting a possible contribution of cold gas through tidal interaction (Fig. \ref{fig:interacting}).

The NGC 2292 – NGC 2293 system is undergoing a close interaction; however, NGC 2292 displays a peculiar  ring morphology aligned with the center of the galaxy, suggesting a distinct interaction history.

The variety of morphologies within our sample illustrates the complex interplay between environmental effects, mergers, and internal galaxy dynamics in shaping the warm gas distribution. A few sources, such as NGC 3923, display optical shells and stellar streams on large scales, indicative of past merger events and interactions. However, similar features are also observed in elliptical galaxies hosting cooling filaments (e.g., M87), where the warm gas is thought to arise from the condensation of the hot halos. Therefore, the presence of extended hot halos in some of our filamentary systems suggests that hot gas cooling may represent an equally important source of warm gas, particularly in galaxies that lack clear signatures of recent mergers.

Finally, an indication of an external origin, such as mergers or interactions with satellite galaxies, is the low gas-phase metallicity, and the similar metallicity of the warm and hot gas phases \citep{olivares2026}. In contrast, gas originating internally from stellar mass loss typically shows super solar metallicities. We leave a detailed analysis of metallicity for a future work as the focus of this study is on the kinematics of the warm gas and the stellar component. 

\subsubsection{Cooling from hot halos}\label{sec:cooling}

\begin{figure*}[ht!]    
    \centering
    \noindent\textbf{Group-scale halos}\\
    \includegraphics[height=0.3\textwidth]{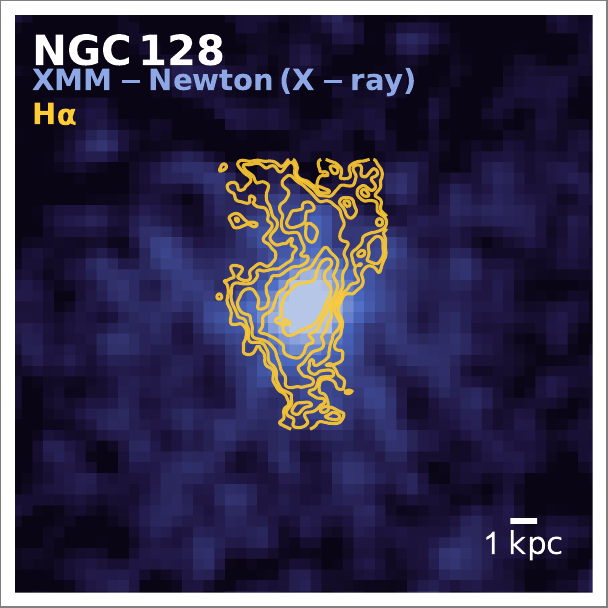}
    \includegraphics[height=0.3\textwidth]{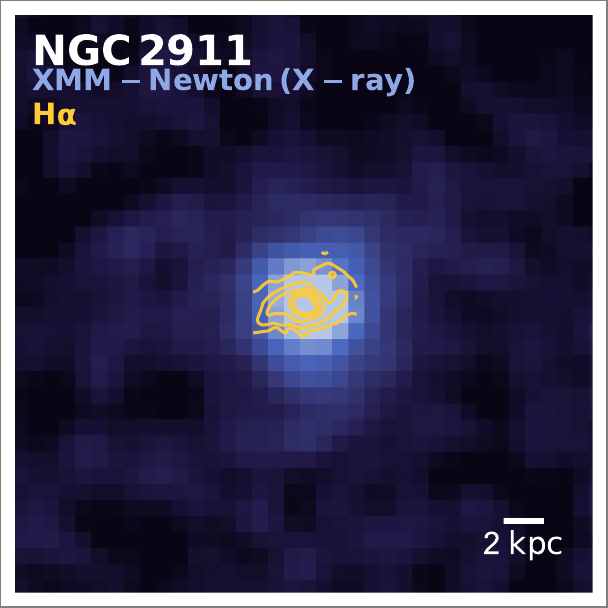}
    \includegraphics[height=0.3\textwidth]{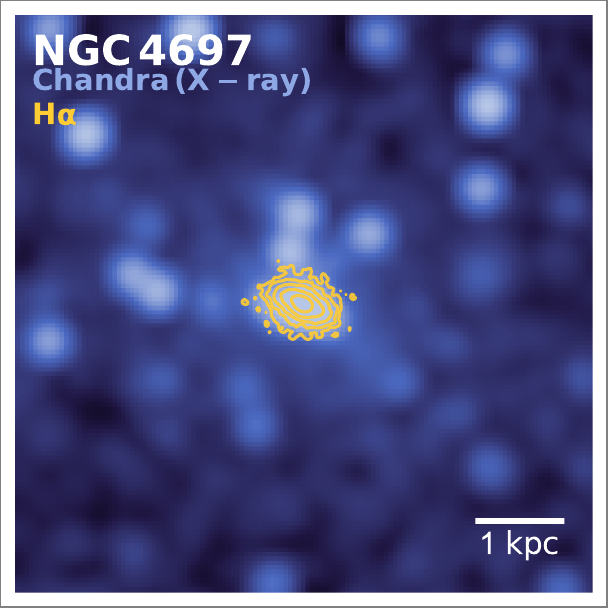}\\
    \includegraphics[height=0.3\textwidth]{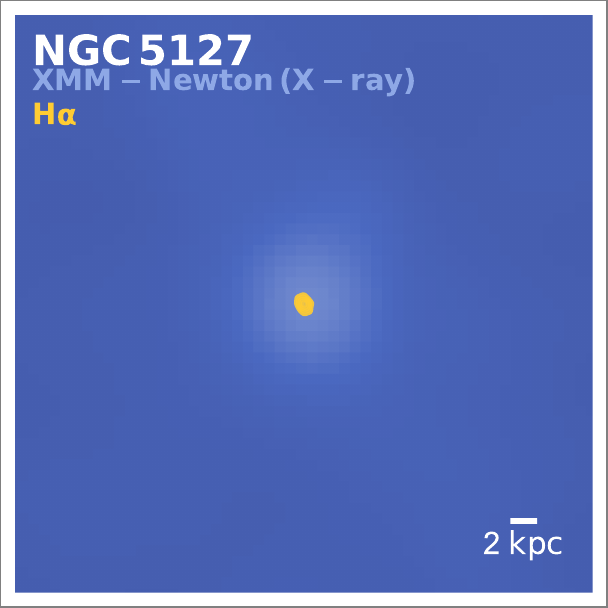}
    \includegraphics[height=0.3\textwidth]{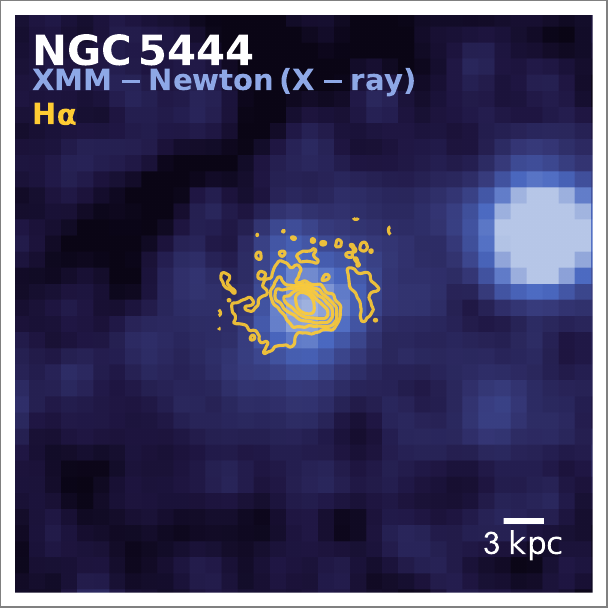}
    \includegraphics[height=0.3\textwidth]{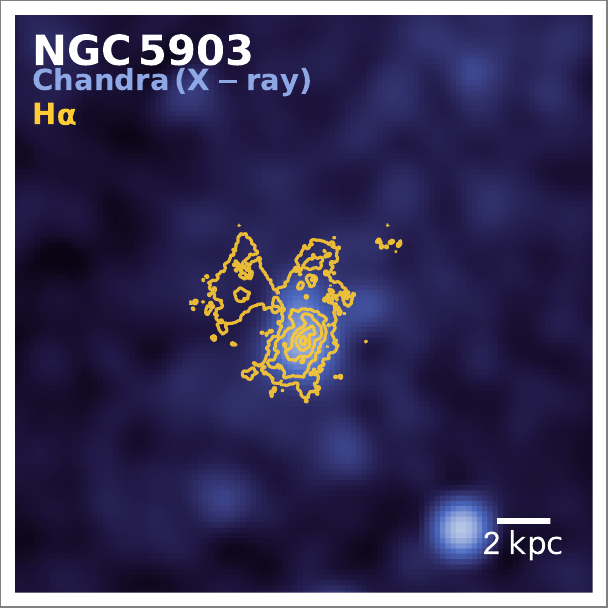}\\
    
    \noindent\textbf{Galaxy-scale halos}\\
    \includegraphics[height=0.3\textwidth]{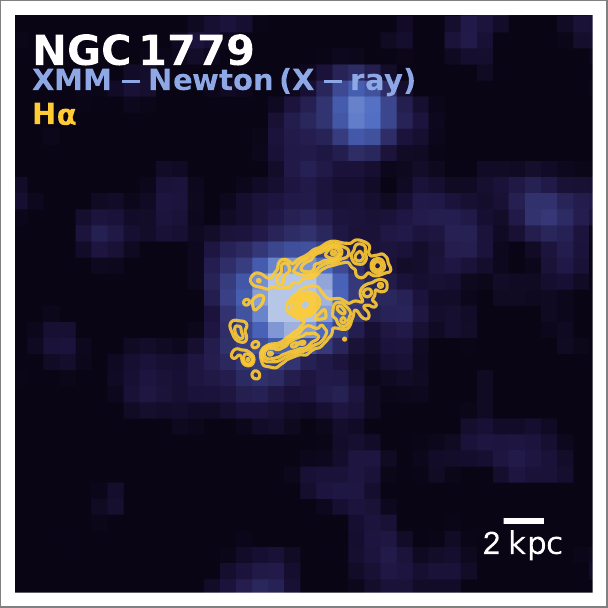}
    \includegraphics[height=0.3\textwidth]{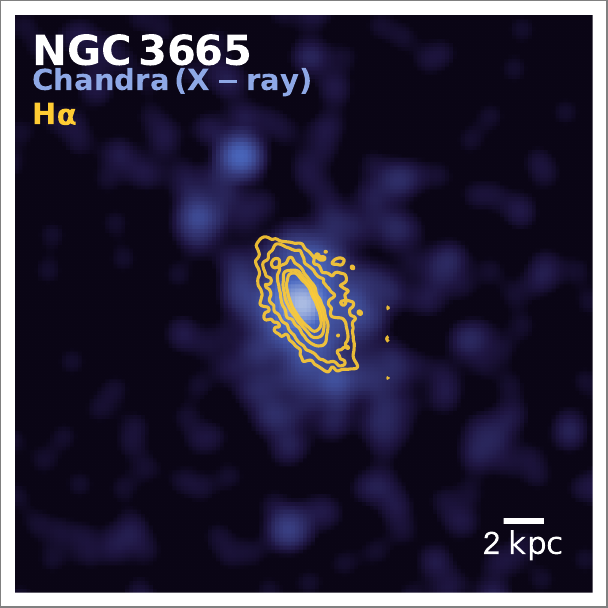}
    \includegraphics[height=0.3\textwidth]{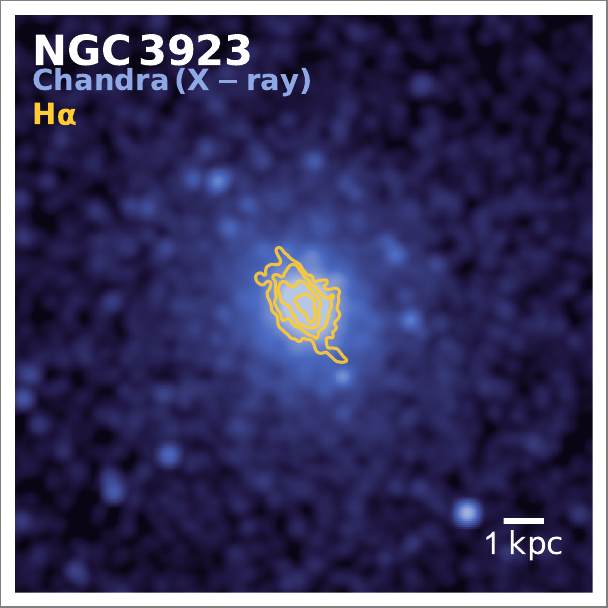}\\
    \includegraphics[height=0.3\textwidth]{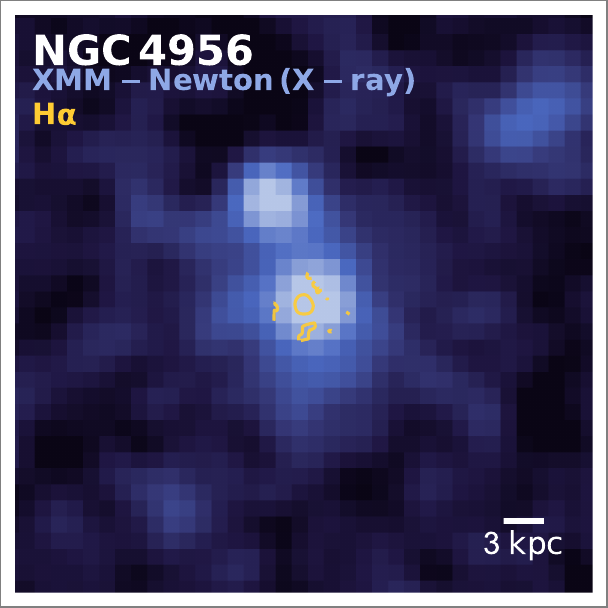}
    \includegraphics[height=0.3\textwidth]{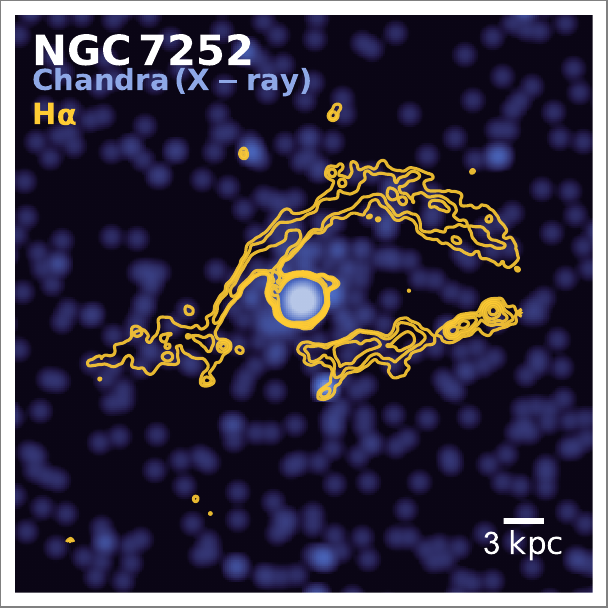}\\
    \caption{X-ray images of our sample from either \textit{XMM-Newton} or \textit{Chandra} observations for group- and galaxy-scale halos (O'Sullivan et al. in prep). Contours from H$\alpha$ are shown with yellow color.}
    \label{fig:Xray_images}
\end{figure*}

Multiphase filamentary structures with complex kinematics have been extensively identified at the centers of cool-core clusters. It is now believed that these filamentary structures are mainly formed through condensation and precipitation of the ICM triggered by AGN feedback \citep{gaspari17,gaspari19,Fournier2024,beckmann22,beckmann22b,Voit2017}. These filaments are typically observed in clusters with strong ICM cooling cores, characterized by low central entropy, usually at 10\,kpc, $K <30\,\rm keV cm^{-2}$, and short central cooling times, $\rm t _{cool} <1\, Gyr$ \citep{Revaz2008,olivares23,cavagnolo2009}. The filaments are often located behind or around the rims of X-ray cavities excavated by radio jets \citep{Olivares2019,russell19,salome08,olivares23,Calzadilla2022}, representing one of the clearest signatures of AGN radio-mode feedback. 

In groups-dominated galaxies selected from the high-richness CLoGS subsamples, studies have found a strong indication of IGrM cooling based on their thermodynamical properties in systems displaying filamentary structures and small disks \citep{Olivares2022,Osullivan2024}. Other examples of well-studied massive groups such as NGC 5044 and NGC 5813, where X-ray cooling, optical filaments, and molecular gas strongly suggest ongoing condensation \citep[e.g.,][]{David2011,Werner2014,Russell2014,osullivan17,Temi2018,Babyk2018,Temi2026}. More broadly, statistical works demonstrate that group-central galaxies in massive groups with low entropy and short cooling time frequently host cold molecular gas, warm ionized nebulae, suggesting that thermally unstable cooling and AGN-driven uplift play a role in fueling multiphase gas in low-mass halos \citep[e.g.,][]{Babyk2018,Lakhchaura2018}. 

In our sample, several galaxies display prominent warm-ionized structures: NGC 128, NGC 3923, NGC 5903, and NGC 7252 host extended filamentary features, while NGC 5061, and NGC 5127 contain compact rotating disks that could trace previous cooling and condensation from their X-ray halos, forming compact, dispersion-dominated structures of low-angular-momentum gas, possibly linked to AGN fueling. It is worth noting that not all filaments necessarily arise from thermally unstable cooling alone. Some systems show clear evidence of recent or ongoing interactions, such as the previously mentioned post-merger galaxy NGC 7252, suggesting that mergers or galaxy–galaxy encounters can also supply and shape the warm gas. Another particularly striking example is NGC 5903, which exhibits an unusual combination of X-ray, radio-continuum, and H$\alpha$ properties and is connected to a giant tidal HI filament \citep{Appleton1990}. As shown by \citet{Appleton1990}, this filament passed close to NGC 5903, and part of the stripped HI appears to be settling into a disk. The H$\alpha$ morphology and kinematics we observe may therefore reflect, at least in part, the accretion of this tidal material rather than solely IGrM cooling. Consistent with this interpretation, our MUSE data reveal a rotating pattern in NGC 5903 that likely traces the settling of this infalling gas.

\begin{figure}[ht!]    
    \centering
    \includegraphics[height=0.9\columnwidth]{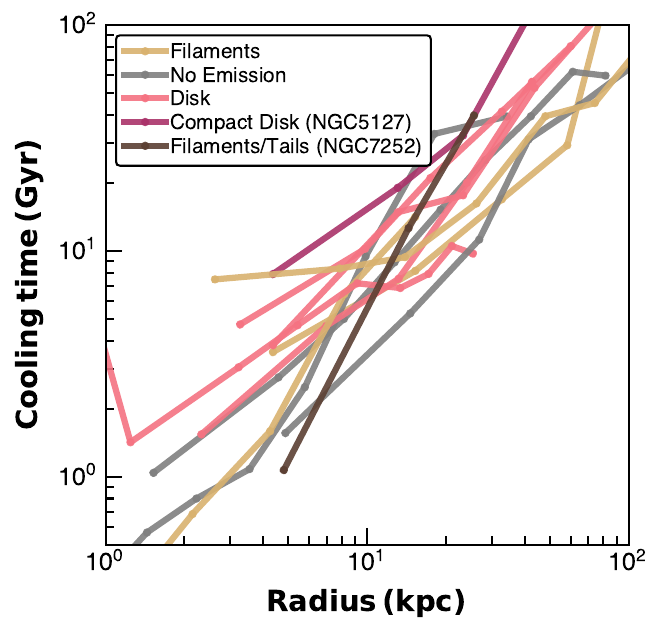}
    \caption{Cooling time profiles for the sources that have detected X-ray halos (O'Sullivan et al. in prep). Filaments sources are shown with light brown lines, while compact and extended disks are displayed with pink and dark pink, respectively. We have also highlighted the filament/tails system: NGC 7252, in brown.}
    \label{fig:cooling_time_profiles}
\end{figure}

X-ray observations of the hot gas are key to unveil the origin of the cold and warm gas in group-dominant galaxies. \textit{Chandra} and \textit{XMM-Newton} observations of our sample detect X-ray halos in most sources (see Table \ref{tab:sample_prop}, O'Sullivan et al. in prep). For example, 9/13 of our system with optical emission lines show group-scale X-ray halos, while 6/13 display galaxy-scale X-ray halos, and and 2/13 have no detected X-ray emission. In figure \ref{fig:Xray_images} we show the distribution of the warm ionized gas together with the hot gas. However, we do not find a correlation between the different categories (disks and filaments), and the properties of the hot X-ray emitting plasma (Fig.~\ref{fig:cooling_time_profiles}). In particular, systems exhibiting extended warm ionized emission do not show systematically shorter central cooling times compared to those without detectable warm gas.  

\begin{figure*}[t]
    \sidecaption
    \includegraphics[width=12cm]{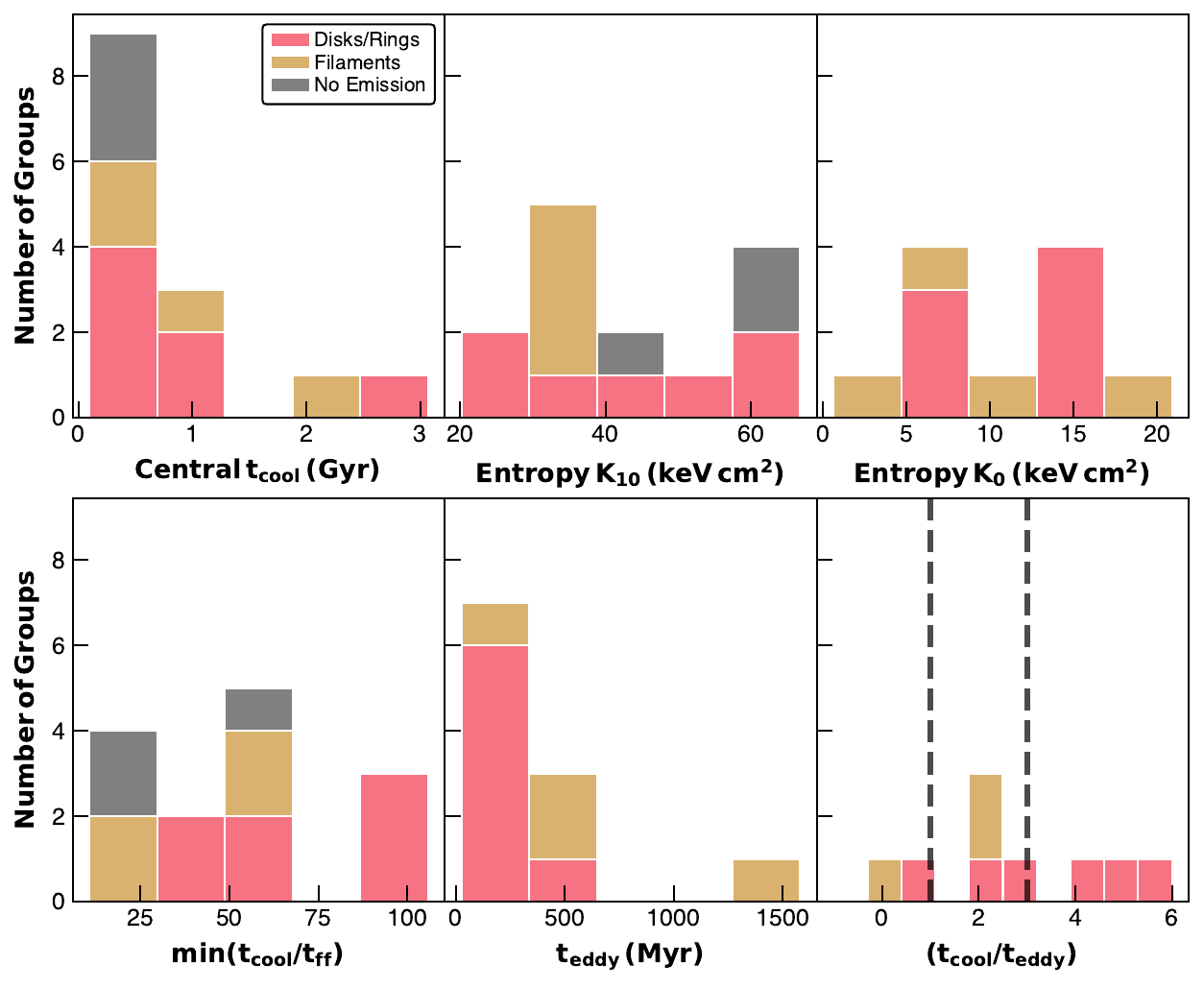}
    \caption{Histograms of selected properties of our sample,
    color-coded by the morphology of the ionized gas:
    filaments (light brown), disks and rings (pink), and
    no detected emission lines (gray).
    Top row: central cooling time, $t_{\rm cool}$, and entropy
    at 1 and 10~kpc, $K_{1}$ and $K_{10}$, respectively.
    Bottom row: minimum cooling-to-free-fall time ratio,
    $\min(t_{\rm cool}/t_{\rm ff})$; eddy turnover timescale,
    $t_{\rm eddy}$; and the ratio of the central cooling time
    to the eddy turnover timescale, $t_{\rm cool}/t_{\rm eddy}$.
    The vertical lines mark $t_{\rm cool}/t_{\rm eddy}=1$ and $3$.}
    \label{fig:histograms}
\end{figure*}

Figure \ref{fig:histograms} show the central entropy values, and central cooling times (see also Table \ref{tab:sample_prop}) for the sources in our sample color-coded by their classification. We have highlighted disks with coupled warm gas and stellar kinematics with diagonal black lines. Contrary to what was found in the dominant galaxies of the high-richness CLoGS subsample using MUSE and \textit{Chandra} observations, in our sample almost all systems, both with and without warm ionized gas, show short central cooling times, $t_{\rm cool}$, and low central entropy values at 1 and 10 kpc, $K_1$ and $K_{10}$, respectively (see Figure~\ref{fig:histograms}). These thermodynamical conditions indicate that many of these halos are in principle prone to cooling and condensation. However, within our sample they are not, by themselves, a reliable predictor of either the presence or the morphology of the warm ionized gas. This suggests that thermodynamic susceptibility to cooling is only one ingredient, while the final gas configuration likely also depends on the level of turbulence and uplift, the angular-momentum content of the cooling material, and the impact of mergers or satellite interactions. In this picture, low-$K$, short $t_{\rm cool}$ systems may still develop relatively settled disks when cooling proceeds with sufficient angular momentum and limited stirring, whereas stronger perturbations are more likely to promote clumpy, filamentary condensation (\citealt{Juranova2019}).

\subsubsection{Cooling versus free-fall and eddy turn-over times}

More details on the condensation process of the hot halos is given by the ratio between the cooling time, $t_{\rm cool}$, and the eddy turn-over timescale, $t_{\rm eddy}$ \citep{gaspari18}. The $t_{\rm eddy}$ corresponds to the time for a turbulent eddy to produce density fluctuations in the hot halo, and is defined as $t_{\rm eddy} = {2 \pi\,r^{1/3}L^{2/3}}/{\sigma_{\rm 3D,los}}$, where $L$ is the turbulence injection scale, $r$ is the radius from the central of the galaxy to the outermost extent of the warm gas, and $\sigma_{\rm 3D,los}$ is the velocity dispersion at the injection scale. The $\sigma_{\rm 3D,los}$ was measured as discussed from the warm gas in Sec\ref{sec:muse}. For the injection scale $L$ we used the size of the warm ionized gas following \citet{Gaspari2018}. When $C$-ratio\,$\equiv t_{\rm cool}/t_{\rm eddy}$ becomes lower than $\sim$\,1-2, filaments and clumpy clouds are expected to condense efficiently from hot halos. As shown in Figure~\ref{fig:histograms}, our sample spans a broad range of eddy turnover times, from  $\sim$30~Myr to 2~Gyr. 
According to the criterion $t_{\rm cool}/t_{\rm eddy} \sim$\,1-2, nearly all filamentary systems are cooling efficiently, with the exception of NGC 5903. In contrast, among the disk/ring systems only NGC 1779 satisfies this condition. In this disk, the stars and gas are dynamically coupled, suggesting that the observed condensation may instead be associated with gas recycled from the underlying stellar population. 

Overall, the warm ionized gas in our sample, in sources where mergers or satellite interactions can be ruled out, likely reflects the combined effects of stellar mass loss and thermally unstable cooling triggered by AGN feedback. A significant fraction of the observed H$\alpha$ emission may originate in numerous small clouds shed by mass-losing post-AGB stars, which are subsequently redistributed by the dynamical influence of radio jets and large-scale motions within the hot halo. It would be useful to clarify that mass loss from AGB stars alone is insufficient to expel gas beyond the galaxy’s gravitational potential, instead, processes such as AGN-driven jets or outflows, or dynamical events like mergers, are required to transport gas to larger scales. In parallel, AGN-driven jets can induce localized compression of the ambient hot gas, promoting radiative cooling and condensation. This mechanism naturally produces the quasi-radial H$\alpha$ filaments frequently observed in group-dominant galaxies, and found in several systems in our sample. Together, these processes support a scenario in which both stellar ejecta and AGN-triggered cooling contribute to the observed multiphase structures.

\subsubsection{Gas-free galaxies}

To place the two coold gas-free galaxies in the broader context of how AGN feeding and feedback regulate the cool gas reservoir, the observed multiphase gas properties are compared with predictions from AGN self-regulation and top-down condensation models to evaluate their consistency across different evolutionary stages. NGC 1395 and NGC 1407 show no evidence of cool gas: deep Spitzer data reveal no significant far-infrared dust emission consistent with a total diffuse dust mass of the order of $\sim 10^4 M_\odot$ \citep{temi09, amblard14}. The H$\alpha$ emission remains undetected, suggesting that these sources may be in a phase where feedback is either inactive, such that cold gas is not produced, or not sufficiently strong to power significant outbursts. 

Alternatively, the AGN in these systems may have undergone a powerful outburst in the past, during which energetic feedback heated and disrupted the cold gas reservoir, thereby suppressing SMBH accretion and subsequent jet activity. In particular, radio observations from GMRT and VLBA of NGC 1407 suggest two cycles of radio outbursts occurring at different epochs \citep{Giacintucci2012}. The current radio activity has released nearly an order of magnitude less energy than that associated with the previous outburst, suggesting that the source has not produced sufficient energy to power a comparable outburst capable of triggering the subsequent formation of colder gas.

\subsubsection{AGN feedback, k-plot}\label{sec:feedback}

\begin{figure*}[ht!]    
    \centering
    \noindent\textbf{Group-scale halos}\\
    \includegraphics[height=0.3\textwidth]{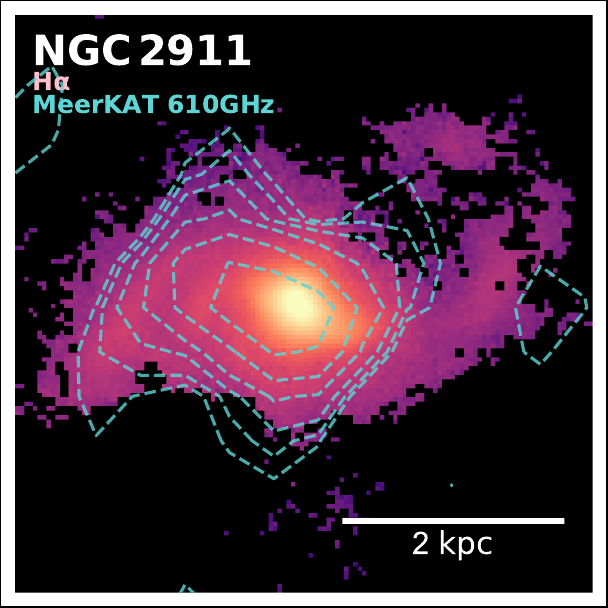}
    \includegraphics[height=0.3\textwidth]{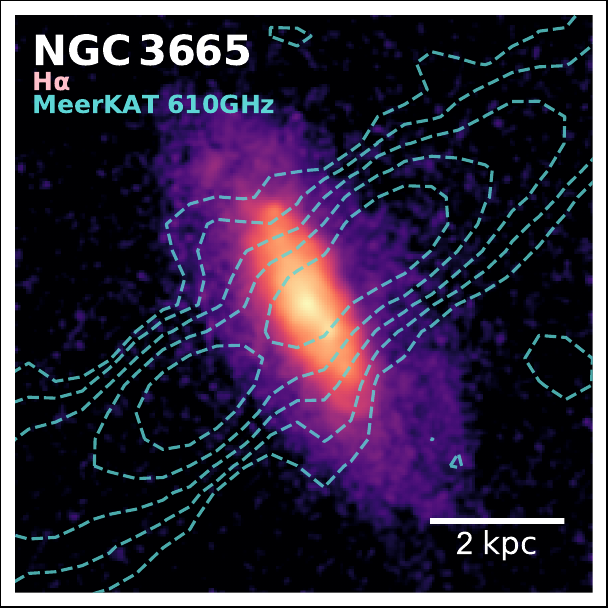}
    \includegraphics[height=0.3\textwidth]{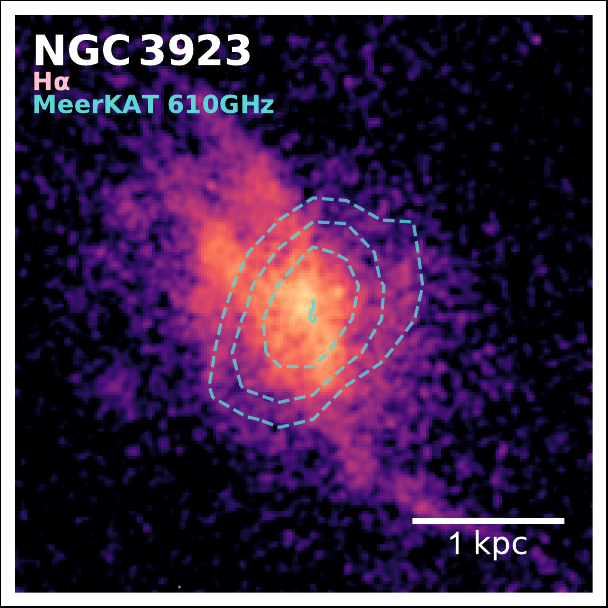}\\
    \includegraphics[height=0.3\textwidth]{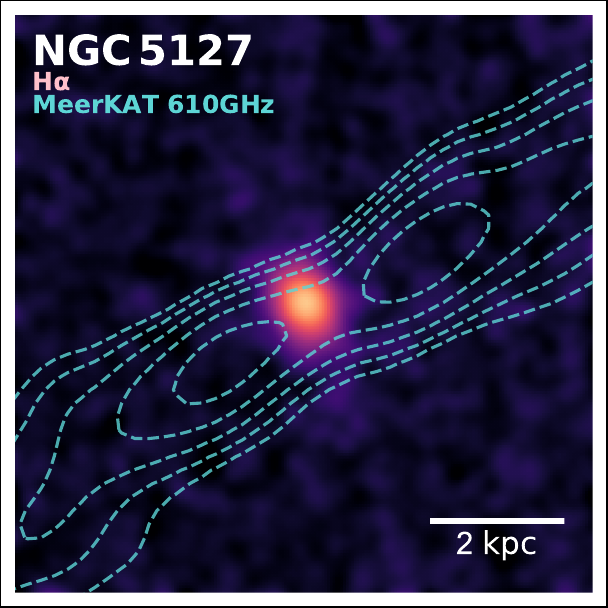}
    \includegraphics[height=0.3\textwidth]{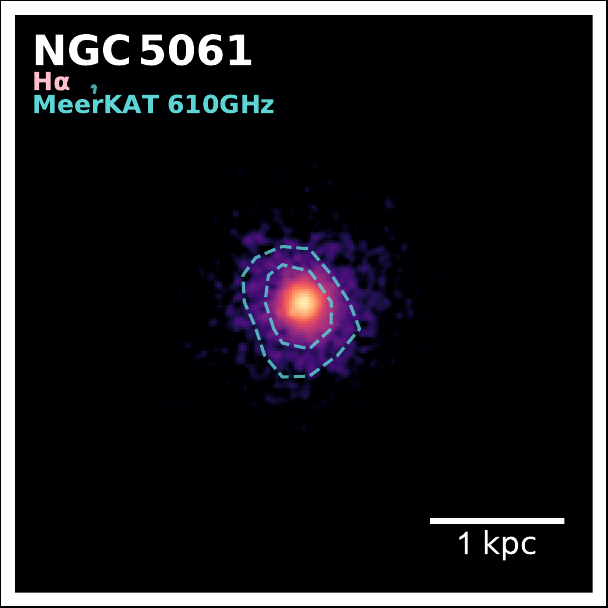}
    \includegraphics[height=0.3\textwidth]{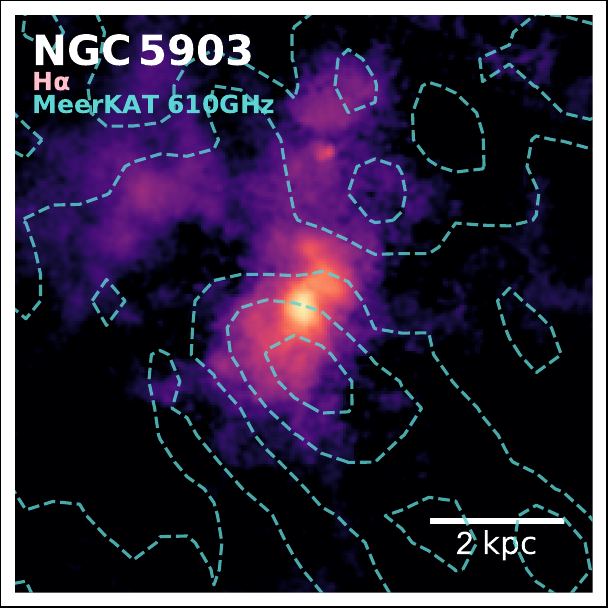}\\
    \caption{H$\alpha$ images for the sources that have extended radio emission. The cyan contours corresponds to the radio 610 MHz emission from the GMRT \citep{Kolokythas2019}. The radio contours are 3$\sigma$, and increase by a factor of 2.
    We note that radio emission of NGC 2199, NGC 3665, and 5127 are dominated by AGN, while NGC 3923 and NGC 5061 have have SFR-predicted radio emission that is $>$50-60\% \citep{Kolokythas2019}. }
    \label{fig:radio_images}
\end{figure*}

Many of our sources may host a radio AGN based on the radio observations. GMRT radio maps reveal a range of morphologies, from compact nuclear sources, diffuse radio remnant jets, large-scale jets extending over several kiloparsecs \citep{Kolokythas2019}, and a few sources which lack radio emission (see Table \ref{tab:sample_prop} for details on each source). We note that the radio emission of some point-like sources have a large contribution ($>$50 - 60\%) by star formation (NGC 128, NGC 1395, NGC 1779, NGC 1395, NGC 3923, NGC 4697, and NGC 5061, NGC 5903, NGC 7252). The radio emission of NGC2911, NGC 5127 and NGC 3665 are AGN dominated.

There are a few sources in our sample that show evidence of AGN feedback, where radio jets appear to interact with the ionized gas traced by H$\alpha$, possibly disturbing, ionizing or redistributing it within the central few kiloparsecs (see Figure \ref{fig:radio_images}). In particular, NGC 2911 (radio emission dominated by AGN) exhibits extended radio emission (cyan contours), as revealed by the 610 MHz GMRT observations \citep{Kolokythas2019}. The radio emission closely follows the direction of the H$\alpha$ filaments and asymmetries, suggesting that the jet may be propagating through an inhomogeneous medium, compressing and ionizing the gas along its path. In NGC 5903, the radio emission (green contours) extends beyond the central region and displays a complex morphology, partially aligned with the filamentary H$\alpha$ emission (see Figure \ref{fig:radio_images}). The H$\alpha$ map reveals asymmetric ionized gas structures that appear to coincide with regions of enhanced radio emission, suggesting that the radio lobes may be interacting with the surrounding interstellar medium. In addition, the southern and northern extensions, where radio emission is stronger, the radio follows the direction of the brightest H$\alpha$ features, and the ionized gas has a higher velocity dispersion. We must caution that a fraction of the radio emission in NGC 5903 is linked to star formation.

Lastly, two sources with gaseous disks, NGC 5127 and NGC 3665, display prominent large- and small-scale radio jets, respectively (see Figure \ref{fig:radio_images}). In both galaxies, the radio emission is spatially aligned along the disk rotation axis of the ionized gas traced by H$\alpha$. Similar features have been observed in other groups and clusters such as Hydra-A and NGC 4261 \citep{olivares22b,rose2019}. The rest of the sources in our sample show point radio (unresolved) sources or lack radio emission.

\begin{figure}
    \centering
    \includegraphics[height=0.5\textwidth]{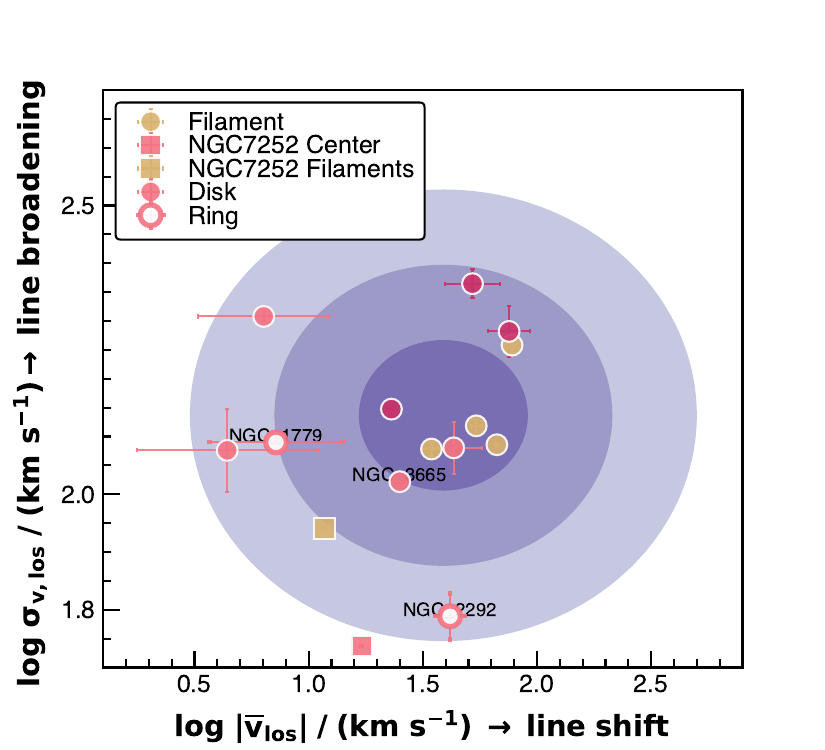}
    \caption{CCA k-plot diagnostic (ensemble beam): line broadening vs. line shift, separating turbulence vs. bulk motions for our sample. The pink points denote the disks, rings, and dark pink shows the compact disks, while the yellow points correspond to the filamentary sources in our sample. NGC 7252, with the central (disk-dominated) region and filamentary regions shown separately with square symbols. We labeled the settled disks and rings NGC 2292, NGC 3665, and NGC 1779.} The purple shaded contours are the 1$\sigma$–3$\sigma$ confidence level predicted via the CCA hydrodynamical simulations \citet{gaspari18}.
    \label{fig:CCA_plot}
\end{figure}

In the current paradigm of self-regulated AGN feedback, gas condensation from the hot halo proceeds in a non-linear, chaotic manner known as CCA or precipitation \citep{gaspari13, gaspari17, Voit2018}. In this framework, turbulent motions and uplift driven by radio jets promote local thermal instabilities in regions where the ratio of cooling to eddy timescale is $t_{\mathrm{cool}}/t_{\mathrm{eddy}}\approx$1--2. In this paradigm, cold clouds and filaments condense out of the hot phase, rain toward the central galaxy, and collide inelastically within the inner few kpc, temporarily boosting the accretion rate onto the supermassive black hole. The resulting feedback outburst heats and stirs the surrounding medium, preventing runaway cooling while sustaining further turbulence and condensation. 

The k-plot provides an ensemble-beam diagnostic of the relative contribution of ordered bulk motions and turbulent line broadening. A dynamically settled, rotation-dominated disk is expected to exhibit relatively narrow intrinsic line profiles, although its global velocity shift may remain small because the approaching and receding sides of the disk can partially cancel when integrated. In contrast, a multiphase medium undergoing strong turbulence, uplift, and chaotic condensation produces a broader distribution of line-of-sight velocities, increasing $\sigma$ relative to the net velocity shift and moving the system toward the low-$|v|$/high-$\sigma$ regime. The k-plot therefore does not replace the spatially resolved morphology and velocity field, but provides a complementary global measure of the balance between ordered and disordered motions. The distribution of our BGGs in this kinematic phase space therefore provides a complementary probe of the interplay between AGN-driven turbulence, cooling-induced condensation, and the re-accretion of multiphase gas in group environments.

We explore this kinematic prediction in Figure \ref{fig:CCA_plot}, where we compare the |$v_{\rm los}$|\,–\,$\sigma_{\rm los}$ distribution for disk and filament dominated systems. The velocity dispersion has been corrected for instrumental broadening. 

Most filamentary systems preferentially occupying the turbulence-dominated region of the diagram, where line broadening exceeds or rivals the net line shift. This is the regime expected when multiphase gas condenses out of a stirred hot halo and is subsequently uplifted, mixed, and chaotically re-accreted \citep[e.g.][]{2020MNRAS.498.4983W}, producing ensemble-beam kinematics dominated by turbulent motions rather than by ordered rotation. 
When considered together with the thermodynamic evidence, in particular the tendency of the filamentary systems to approach the raining threshold $t_{\rm cool}/t_{\rm eddy}\sim 1$--2, the k-plot therefore strengthens the interpretation that a substantial fraction of the warm ionized gas in these BGGs is linked to a CCA-like condensation cycle. However, the distribution of the disk-dominated systems is more heterogeneous, indicating that the k-plot alone does not uniquely identify the origin of the gas. The ordered disk systems likely trace a different balance between cooling, turbulence, and angular momentum.

\begin{table*}[htb]
\caption{Summary of the multiphase gas properties.}
\centering
\footnotesize
\begin{tabular}{lccccccc}
    \noalign{\smallskip} \hline \hline \noalign{\smallskip}
    BGG & Morphology & Stellar dynamics & X-ray  &  Radio & $t_{\rm cool}/t_{\rm eddy}$ & $\rm M_{\rm SML}/M_{\rm gas}$ & Notes \\
    \noalign{\smallskip} \hline \noalign{\smallskip}
    {\it Filaments}   &  &&&&\\
    
  \small  NGC 128 & filaments & decoupled   & hot halo & & $\approx1$ & $>>1$ & cooling \\
  \small  &  & (interacting) & & &     \\

  \small  NGC 3923 & filaments & decoupled  & hot halo & &  $\approx1$ &  $\rm >>1$ & cooling\\
  \small   &  &   & & &  \\
  
  \small  NGC 5903 & filaments & decoupled  & hot halo  & radio-warm gas &  $>> 1$ & $<<1$  & tidal interactions, \\
  \small   &  &   &  & interaction &  & &past interaction/merger \\
  
  \small  NGC 7252 & filaments  & coupled  & hot halo & & $\approx1$ & $\rm <<1$ & past interaction/merger,\\
  \small   &  tails/inner ring & KDC  (post-merger) &  & & &&   some cooling \\
\hline

    {\it Disks}   &  \\
  \small  NGC 1779 & inner disk & coupled (KDC) & hot halo & &$\approx1$ & $\rm <<1$ & past merger/accretion \\
  \small   & + outer ring &  & & & & &  some cooling\\

  \small NGC 2292 & ring & coupled  & no halo & &  & $\rm  \approx1$  & past merger/accretion\\
     &  &  & & &  \\

  \small  NGC 2911 & disk + tails & decoupled  & hot halo & radio-warm gas & $\approx2$ & $\rm <<1$ & merger/accretion\\
  \small   &  &   & &  interaction &  & & some cooling? \\

  \small  NGC 3665 & disk & coupled  & hot halo & large-scale jets &  $\approx1$ & $\approx1$ & internal \\ 
  \small   &  &   &  &  & & & + some cooling \\ 

 \small   NGC 4697 &  disk + diffuse gas & coupled  & hot halo  & & $>>1$ & & past merger/accretion\\
 \small   &  &   &  & &  \\

  \small  NGC 4956 & disk + tails & coupled  & hot halo & &  $>>1$ & $ \approx1$ & past merger/accretion\\
\small   &  &   & & &  & & + some internal\\

 \small   NGC 5061 & compact disk & coupled (KDC) & no halo & & & $>>1$ & residual cooling?\\
  \small    &  &   & & &  \\

 \small   NGC 5127 & compact disk & decoupled  & hot halo & large-scale jets & $>>1$ &$\rm <<1$ & residual cooling?  \\
  \small    & &   &  &  &   \\

 \small   NGC 5444 & disk + diffuse gas & decoupled & hot halo & & $>>1$ & $\rm >>1$ & past merger/accretion\\
 \small    &  arms &  & & &  \\

    \noalign{\smallskip} \hline \noalign{\smallskip}
\end{tabular}
\tablefoot{KDC: kinematically decoupled core.\\
$t_{\rm cool}/t_{\rm eddy}$: Ratio between cooling time and eddy turnover timescale. \\
$M_{\rm SML}/M_{\rm gas}$: Ratio between rxpected gas from stellar mass loss over total cold and warm gas mass. \\

}
\label{tab:summary_origin}
\end{table*}

\section{Conclusions}

This work presents new MUSE observations of the dominant galaxies in low-mass galaxy groups selected from the low-richness subsample of the CLoGS survey. We observed 17 systems with MUSE, out of the 27 galaxies included in this subsample. We emphasize that the robust outcomes of this work are the morphology fractions, the statistics of kinematic coupling, and the multiwavelength associations. Below we summarize our main findings. 
\begin{enumerate}
    \item We detected warm ionized gas in 77\% (13 out of 17) of the observed systems. The morphology and kinematics of the ionized gas are diverse exhibiting filamentary distribution, extended and compact disks and rings. Among these, 9 galaxies show gas distributions and velocity fields consistent with ordered rotation. Some of those disks also displayed tails and diffuse structures. 4 systems show disturbed kinematics and irregular morphologies consistent with filaments.

    \item The kinematics of the stars is consistent with regular rotation in most galaxies of our sample (11/13), while only two have non regular rotation (2/13). This is contrary to what is found at the center of massive clusters where most dominant galaxies are consistent with random motions \citep{Olivares2019}. Also, 3/11 regular rotating galaxies have kinematically decoupled cores (KDC) which indicated a past merger or accretion event.

    \item We compared the position angles ($PA$) of the stellar and gaseous components and found that 6 rotating gaseous disks and rings exhibit coupled kinematics, with $\Delta PA < 20\degr$, that may suggest some contribution from stellar mass loss. However, even 4 kinematically-aligned systems ($\Delta PA \approx 0^\circ$) show tails, diffuse gas, or outer rings, suggesting a previous accretion event. One additional source, NGC 7252 (filaments), also shows $\Delta PA < 20\degr$, although its overall gas velocity field does not follow the stellar rotation pattern.

    \item We find that 3/9 disks exhibit decoupled gas and stellar kinematics ($\Delta PA > 20^\circ$), indicating an external origin for their gas (e.g., mergers or accretion). Ring-like structures further support an external origin, likely tracing past interactions. In contrast, compact disks are likely consistent with centrally condensed gas, possibly linked to residual cooling and AGN fueling.

    \item Almost all galaxies with filamentary or extended gaseous structures, and not symmetric rotation, exhibit decoupled kinematics between gas and stars, with $\Delta PA > 20\degr$. Some of these systems, such as NGC 128, which is interacting with a nearby companion, and NGC 7252, a well-known major merger remnant, suggest that a fraction of their cold gas may have been acquired through interactions or mergers. Also NGC 5903 could have acquired cold gas through a previous merger or interaction. Despite that, most of the filamentary systems (with the exception of NGC 5903), have thermodynamical properties consistent with gas cooling and condensation from their surrounding hot halos, which suggests some contribution through that channel.

    \item The clearest signatures of a CCA-like condensation regime are found in the filamentary systems, which preferentially combine disturbed, turbulence-dominated kinematics with $t_{\rm cool}/t_{\rm eddy}\sim 1-2$. In contrast, the rotation-dominated and compact-disk systems appear more heterogeneous: a few lie outside the CCA locus and/or show elevated $C$-ratios, consistent with more settled, angular-momentum-supported gas linked to recycling or past accretion, while others may still retain some contribution from residual cooling. Overall, our results favor CCA as an important channel for the filamentary population, but indicate that the disk population likely reflects a broader mix of internal recycling, external accretion, and cooling-driven condensation.

\end{enumerate}

Rather than  separating into discrete classes, the warm ionized gas morphologies and kinematics in group-dominant galaxies appear to define a continuum, shaped by the varying interplay of stellar recycling, external accretion, and cooling from the hot halo. In this picture, low-richness BGGs occupy a transitional regime between isolated ellipticals and cluster-dominant galaxies: some systems retain relatively settled, angular-momentum-supported gas, while others develop disturbed filamentary structures linked to condensation and feedback. The diversity observed in our sample therefore suggests that no single channel dominates universally; instead, the origin and fate of the multiphase gas are regulated by the balance between internal supply, environmental accretion, halo thermodynamics, and the onset of CCA-like precipitation in the more strongly perturbed systems.

\begin{acknowledgements}
VO acknowledges support from the DICYT ESO-Chile Comite Mixto PS 1757, FONDECYT Regular 1251702, and DICYT AYUDANTE\_DICYT 541, Vicerrectoría de Investigación, Innovación y Creación.

MG acknowledges support from the ERC Consolidator Grant \textit{BlackHoleWeather} (101086804).
FU acknowledges the financial contribution from contract PRIN 2022 - CUP J53D23001610006.
MA is supported by FONDECYT grant number 1252054, and gratefully acknowledges support from ANID Basal Project FB210003,  ANID MILENIO NCN2024\_112 and ANID + Vinculación Internacional + FOVI250261. 
KK acknowledges funding support from the South African Radio Astronomy Observatory (SARAO) and the National Research Foundation (NRF; grant UID: 97930).
ET gratefully acknowledges financial support from ANID CATA-BASAL - FB210003, and FONDECYT Regular - 1241005 and 1250821.

Based on observations collected at the European Organisation for Astronomical Research in the Southern Hemisphere under ESO programme(s): 0110.A-4397(A), 0112.A-2375(A), and 099.B-0281(A)

\end{acknowledgements}


% \begin{thebibliography}{}
\bibliographystyle{aa}
\bibliography{references_all}

\begin{appendix} %First appendix
\section{Distribution and kinematics of the ionized gas}\label{sec:fullsample}

\subsection{Filamentary sources}

NGC 128 - this galaxy presents an elongated morphology, extending in the N direction potentially produce by the interaction with NGC 127. The gas velocity ranges between -120 and 160 km s$^{-1}$, with the velocity dispersion ranging between 60 and 250 km s$^{-1}$.\\
\\
NGC 3923 has been reported as a shell galaxy, likely formed as the result of a past merger \citep{1986A&A...166...53D,2012MNRAS.421.1485N}. It exhibits filamentary gas structures extending along the NE and SW directions. The NE filaments reach line-of-sight velocities of $\sim200$ km s$^{-1}$, while the SW filaments reach velocities of $\sim-200$ km s$^{-1}$. The velocity dispersion in the central region spans a wide range, from $\sim50$ to $\sim500$ km s$^{-1}$.\\
\\
NGC 5903 - this galaxy has been described in \cite{2018MNRAS.473.5248O}, as one of the dominant galaxies of a galaxy group, along with NGC 5898. It shows a long filament at the NE. Gas velocities range between -200 and 200 km s$^{-1}$. The velocity dispersion ranges between 40 and 300 km s$^{-1}$. \\
\\
NGC 7252 - result of a collision of two gas-rich disc galaxies \citep{2018A&A...614A..32W}. It shows an inner gas disc with two tidal tails. The gas velocity ranges between -170 and 170 km s$^{-1}$, and the velocity dispersion reaches up to 170 km s$^{-1}$.

\begin{figure*}[ht!]
 \textbf{Filaments}\\
    \centering
    \includegraphics[height=0.65\textwidth]{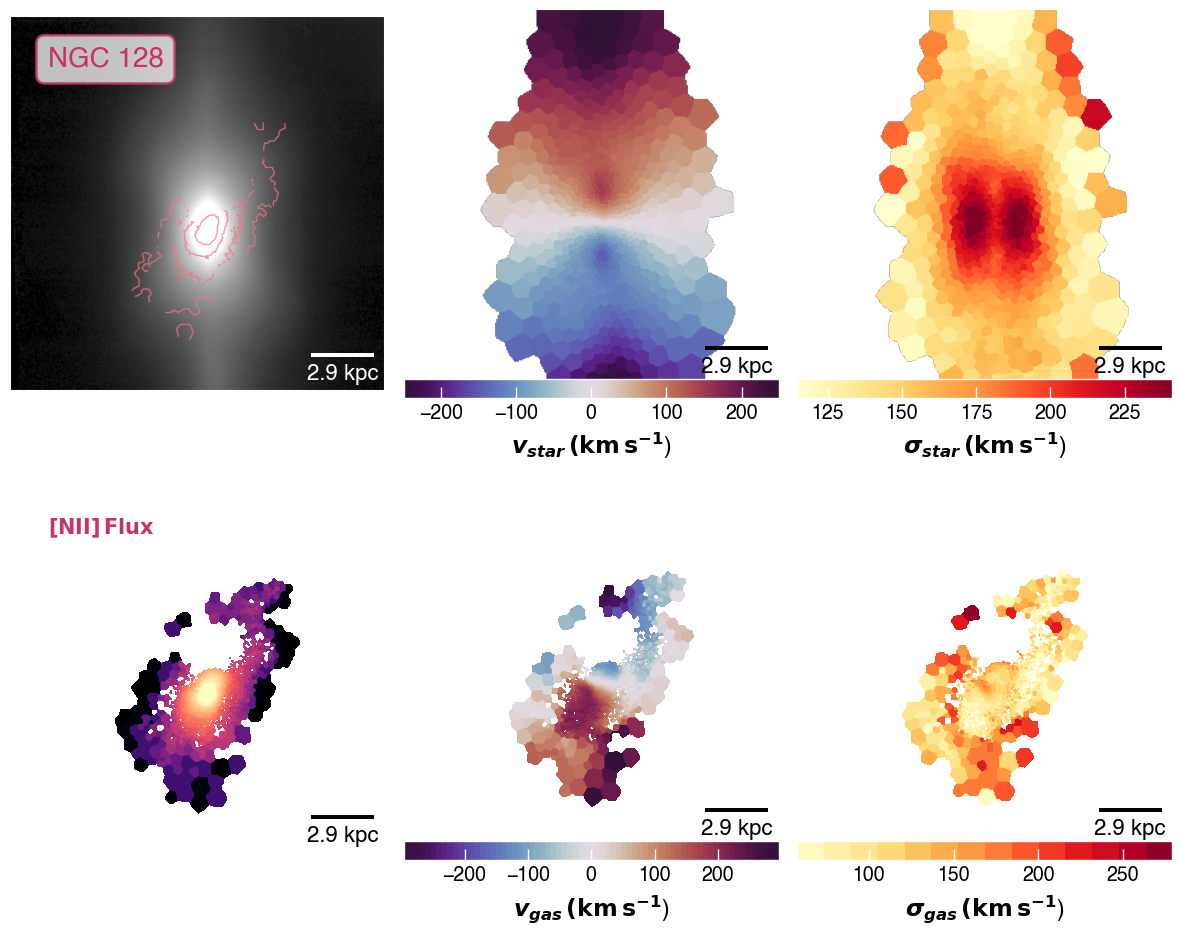}
    \includegraphics[height=0.65\textwidth]{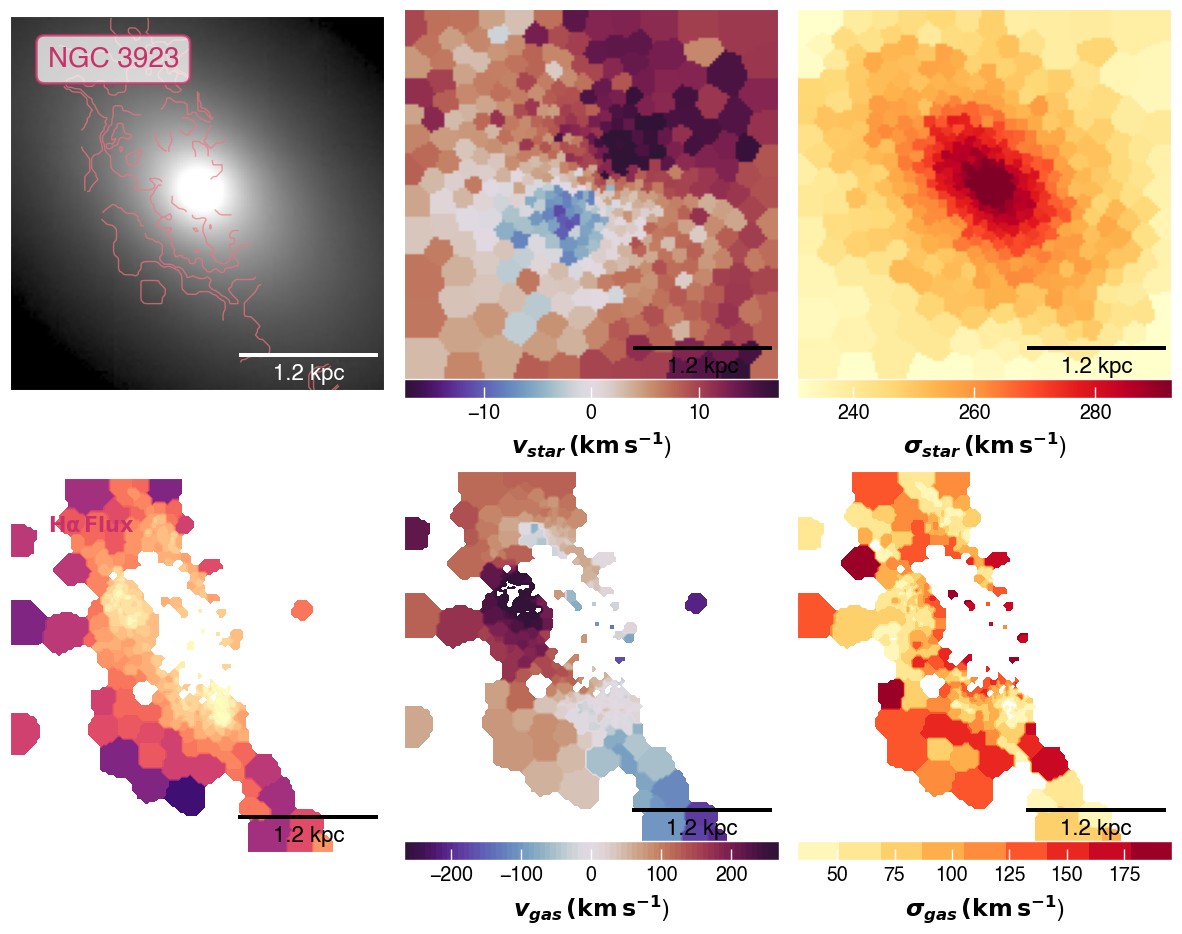}
    \caption{Stellar and warm ionized gas distribution and kinematics for our MUSE sample. The upper panels show the stellar maps derived from the optical continuum between 5400–5600 Å: flux distribution (left), velocity (middle), and velocity dispersion (right). Green contours indicate the H$\alpha$ emission distribution. The lower panels show the corresponding warm ionized gas maps: H$\alpha$ or [NII] maps flux distribution (left), velocity (middle), and velocity dispersion (right).}

    \label{fig:Halpha_distribution}
\end{figure*}

\begin{figure*}[ht!]    
    \centering
    \includegraphics[height=0.65\textwidth]{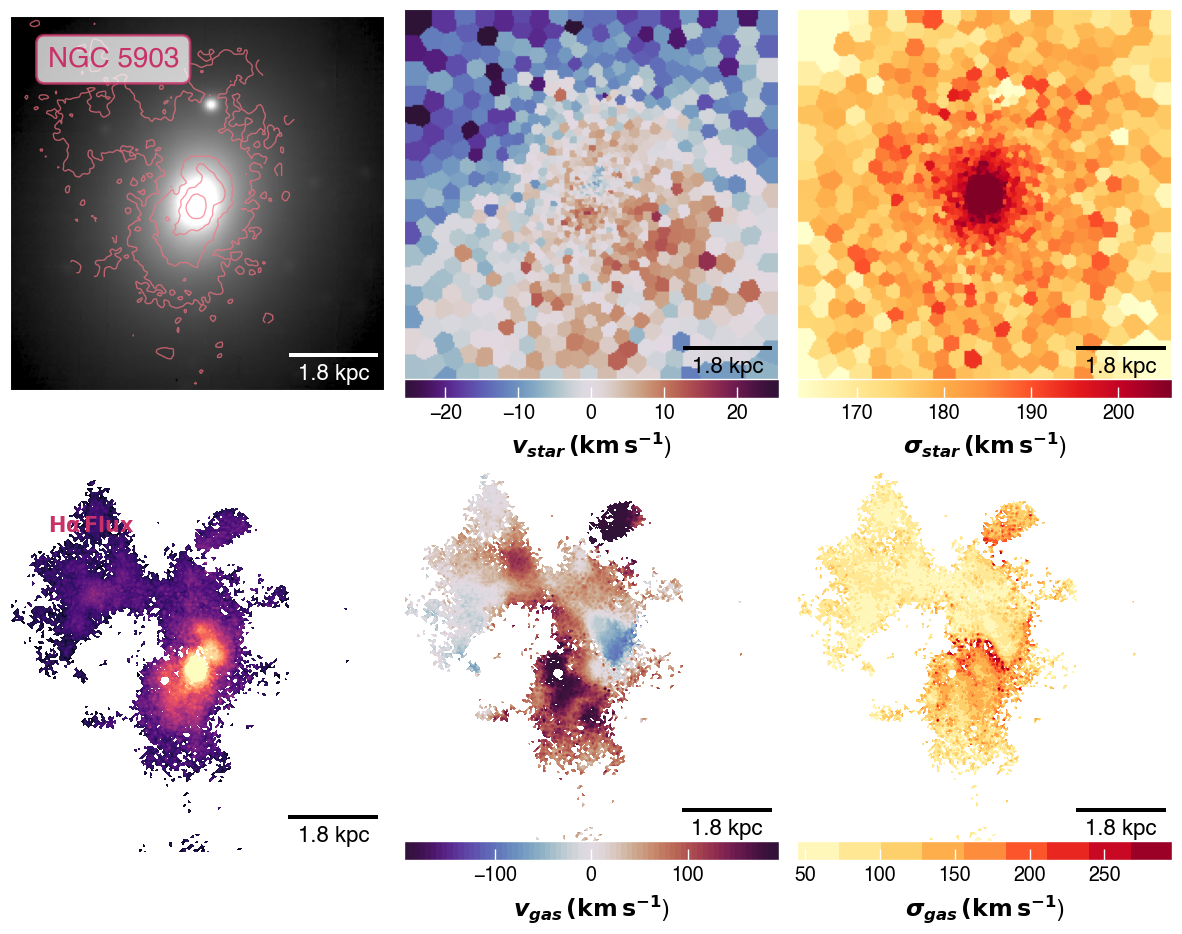}
    \includegraphics[height=0.65\textwidth]{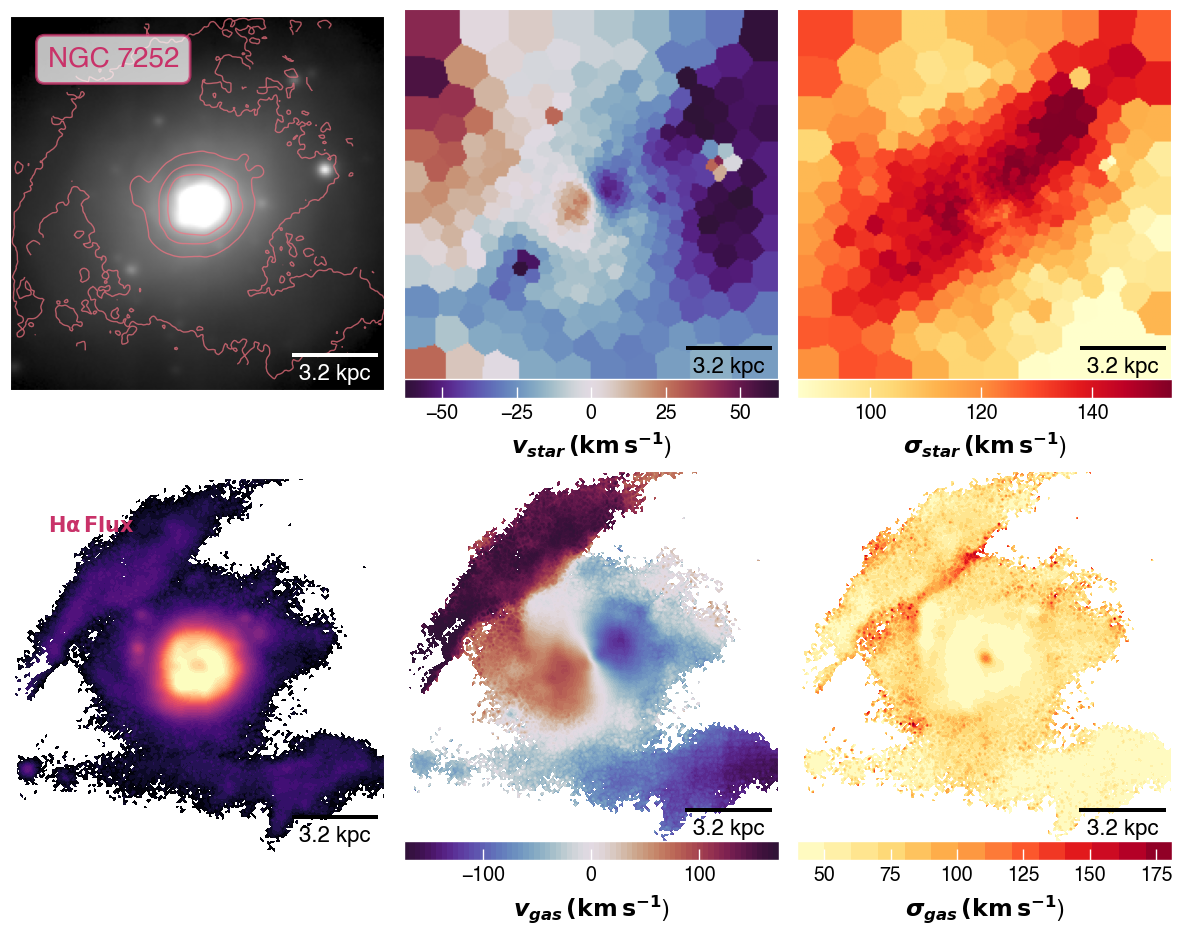}

    \label{fig:ngc5903_mosaic}
    \label{fig:Halpha_distribution2}
\end{figure*}

\subsection{Compact and extended rotating disks}

NGC 1779 — This galaxy presents several clumps located near the disk. The gas velocity ranges between $-250$ and $240$ km s$^{-1}$, with velocity dispersions between $40$ and $150$ km s$^{-1}$.\\
\\
NGC 2911 — This galaxy shows filamentary structures extending in the N and S directions. The gas velocity ranges between $-230$ and $250$ km s$^{-1}$, with velocity dispersions between $50$ and $230$ km s$^{-1}$.\\
\\
NGC 2292 — Part of the galaxy pair NGC 2292/NGC 2293 \citep{2005A&A...443..373S}. It shows a disk-like morphology with a central region lacking emission in some lines (e.g., H$\alpha$, [\textsc{N II}]$\lambda6584$). The gas velocity ranges between $-200$ and $200$ km s$^{-1}$, and the velocity dispersion between $40$ and $130$ km s$^{-1}$.\\
\\
NGC 3665 — This galaxy hosts a weak double-jet radio source oriented perpendicular to the disk plane. The gas velocity ranges between $-300$ and $300$ km s$^{-1}$, with velocity dispersions between $40$ and $150$ km s$^{-1}$.\\
\\
NGC 4697 — The brightest member of the poor NGC 4697 group, which contains five additional lower-mass members \citep{2017MNRAS.468.4675D}. The gas velocity ranges between $-200$ and $200$ km s$^{-1}$ while the velocity dispersion spans from $40$ to $500$ km s$^{-1}$.\\
\\
NGC 4956 - this galaxy has been classified as a S0 galaxy \citep{2012MNRAS.421.1485N}. It presents two small filaments pointing toward E and W. The gas velocity ranges between -200 and 200 km/s, with velocity dispersions ranging between 60 and 400 km s$^{-1}$.\\
\\
NGC 5061 — This galaxy exhibits a compact gas distribution, with velocities between $-150$ and $150$ km s$^{-1}$ and velocity dispersions ranging from $130$ to $400$ km s$^{-1}$.\\
\\
NGC 5127 — This galaxy hosts the radio source B2 1321+31. The gas velocity ranges between $-100$ and $100$ km s$^{-1}$, and the velocity dispersion between $50$ and $160$ km s$^{-1}$.\\
\\
NGC 5444 — This galaxy shows gas velocities between $-250$ and $250$ km s$^{-1}$ and velocity dispersions between $40$ and $200$ km s$^{-1}$.\\

\begin{figure*}[ht!]    
    \centering
    \textbf{{Compact rotating disk}}\par\medskip    

    \includegraphics[height=0.6\textwidth]{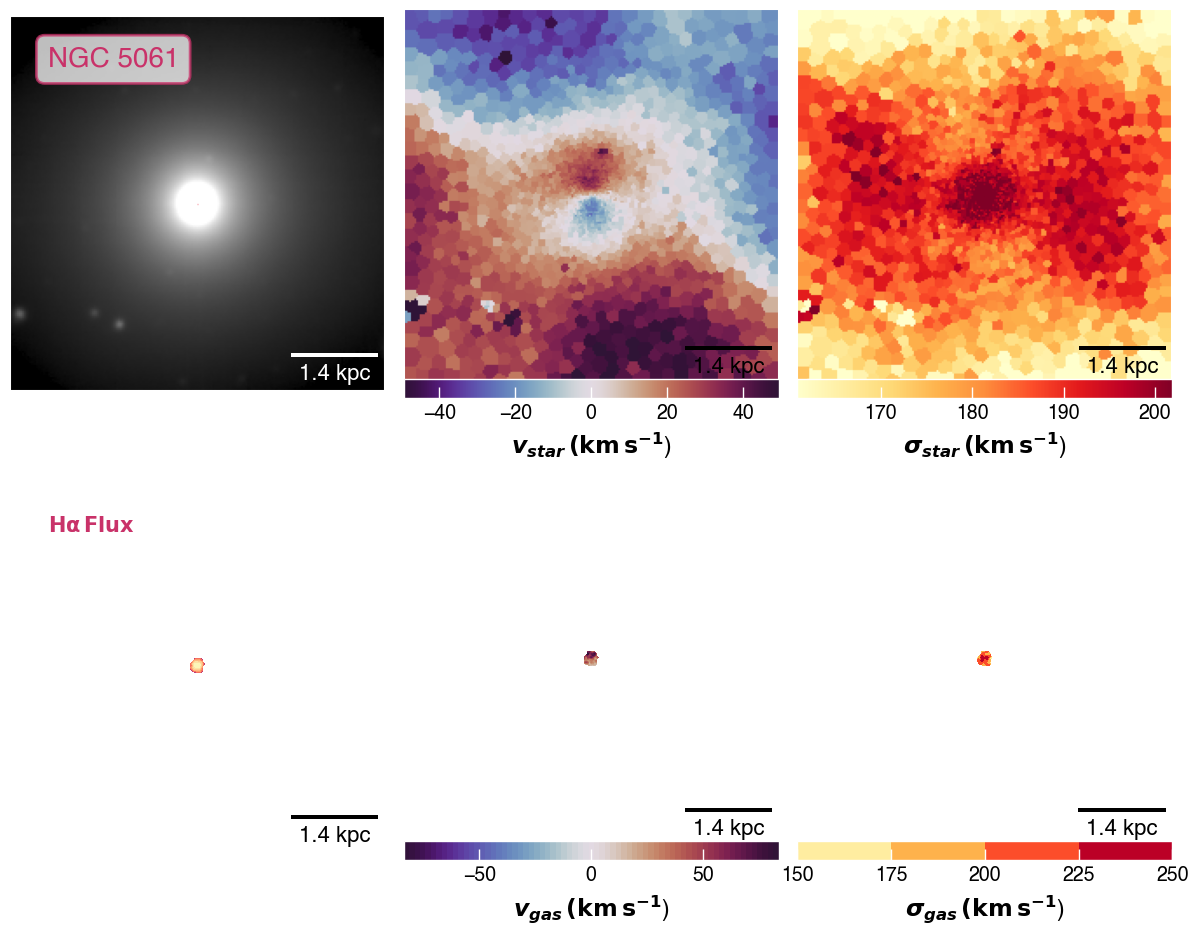}

    \includegraphics[height=0.6\textwidth]{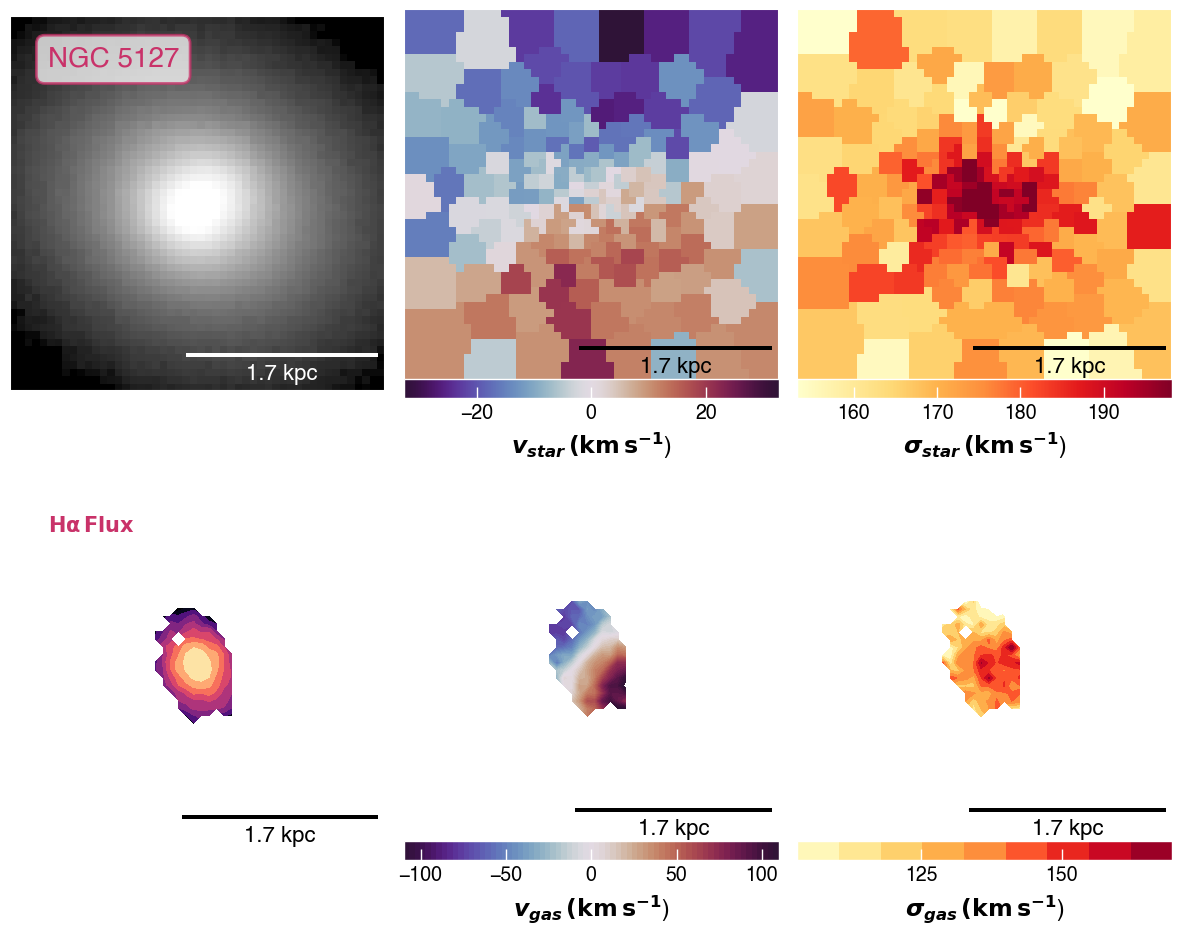}

    \caption{Same as Fig.\ref{fig:gas_distribution}}
    \label{fig:Halpha_distribution3}
\end{figure*}

\begin{figure*}[ht!]
    \raggedright
    \textbf{{Extended rotating disk}}\par\medskip    
    \centering
    \includegraphics[scale=0.5]{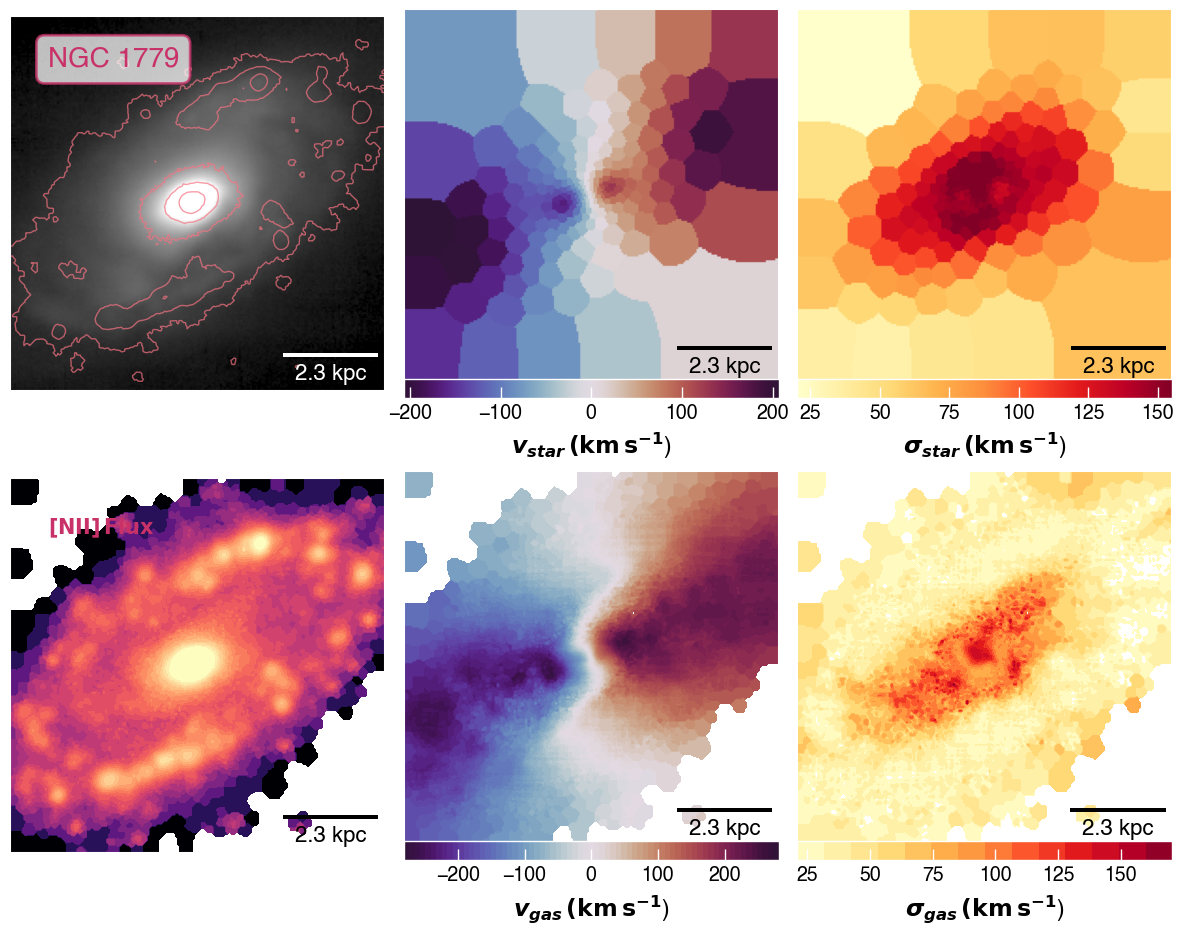}
    \includegraphics[height=0.65\textwidth]{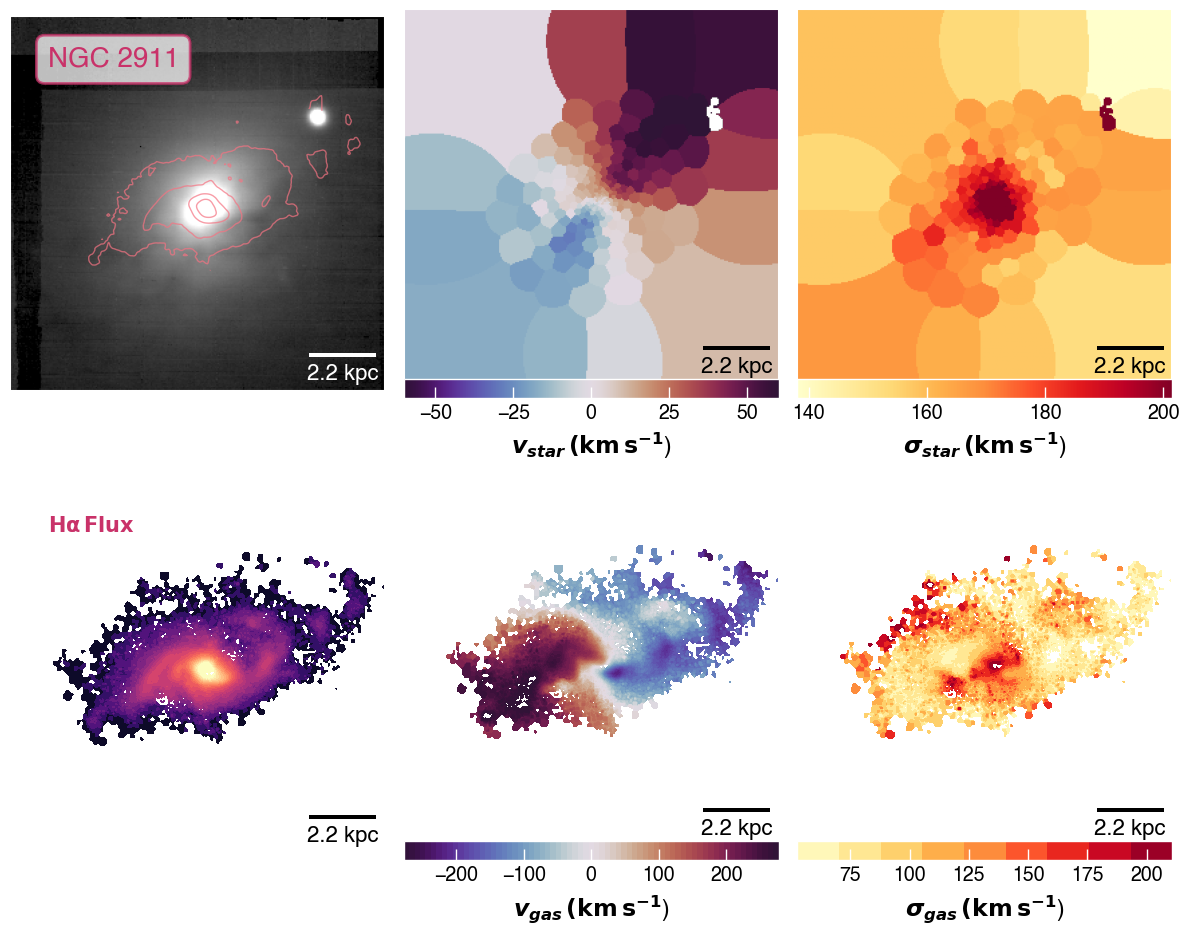}

    \caption{Same as Fig.\ref{fig:gas_distribution}}
    \label{fig:Halpha_distribution4}
\end{figure*}

\begin{figure*}[ht!]    
    \centering
    \includegraphics[height=0.65\textwidth]{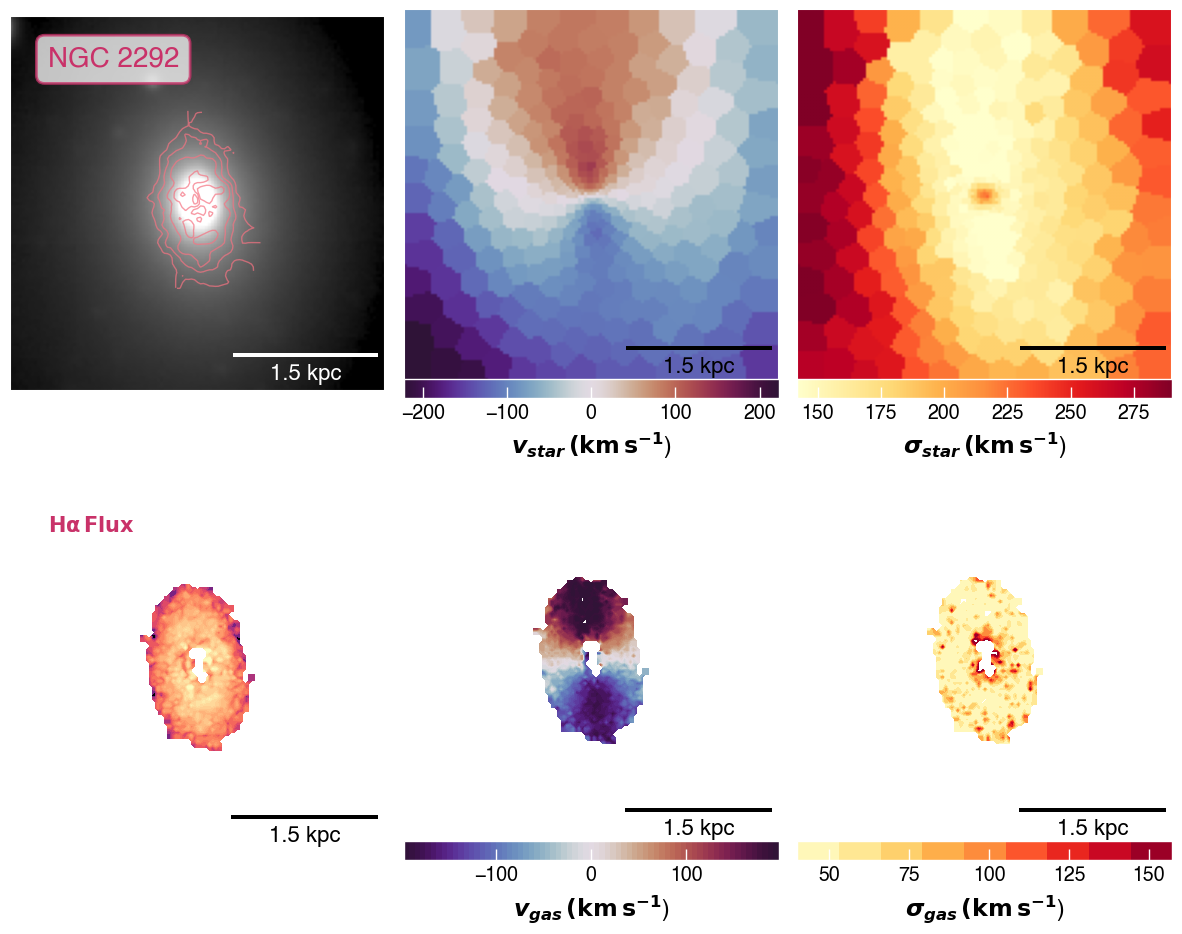}

    \includegraphics[height=0.65\textwidth]{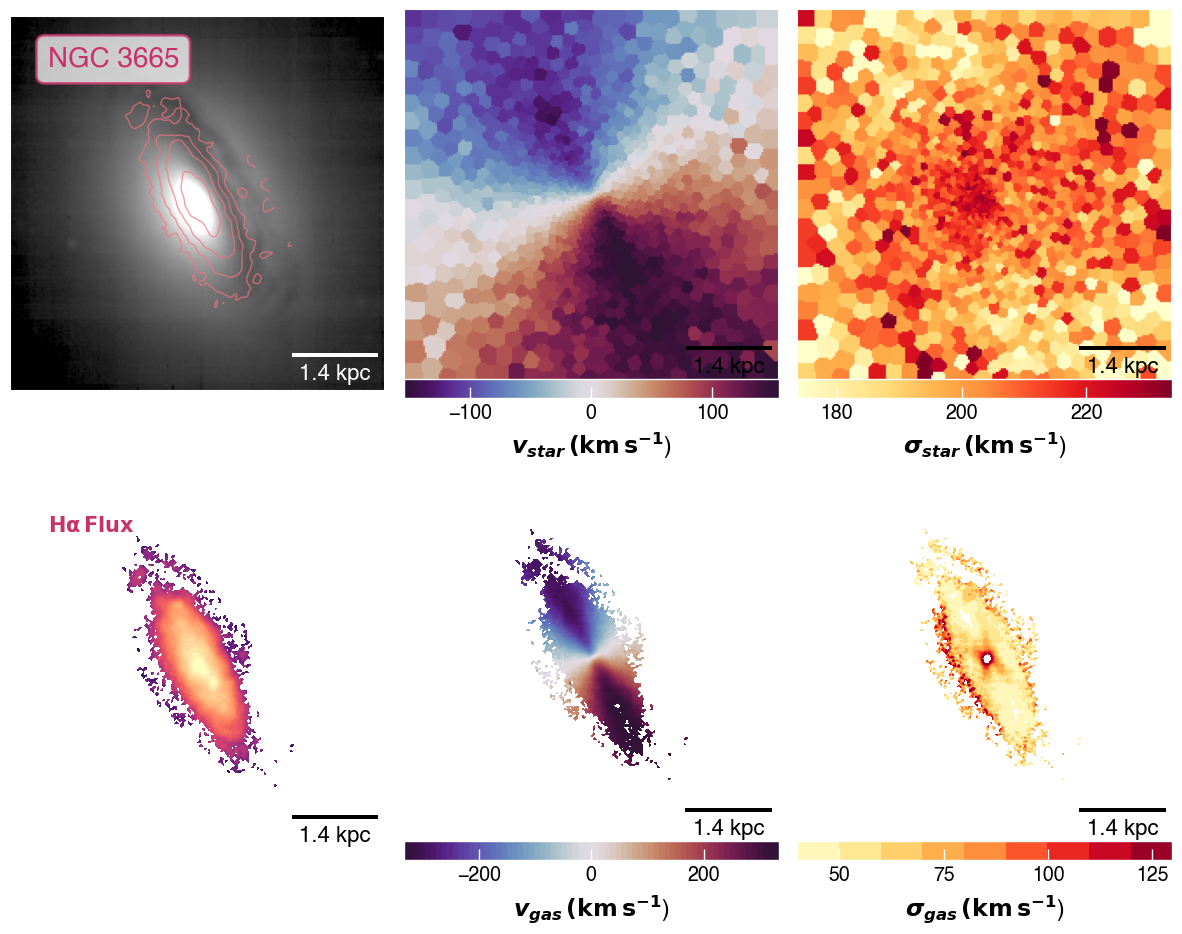}

    \caption{Same as Fig.\ref{fig:gas_distribution}}
    \label{fig:Halpha_distribution5}
\end{figure*}

\begin{figure*}[ht!]    
    \centering
    \includegraphics[height=0.65\textwidth]{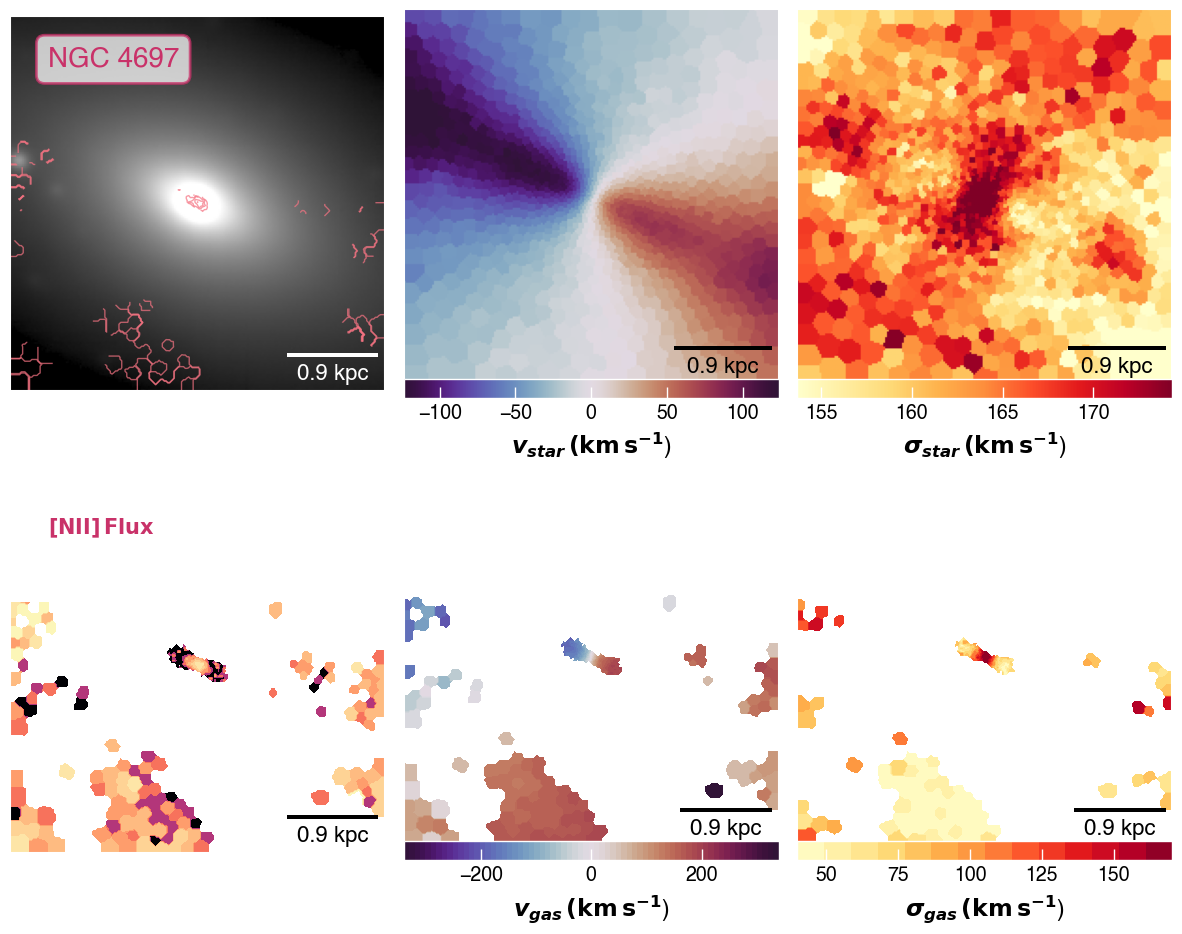}

    \includegraphics[height=0.65\textwidth]{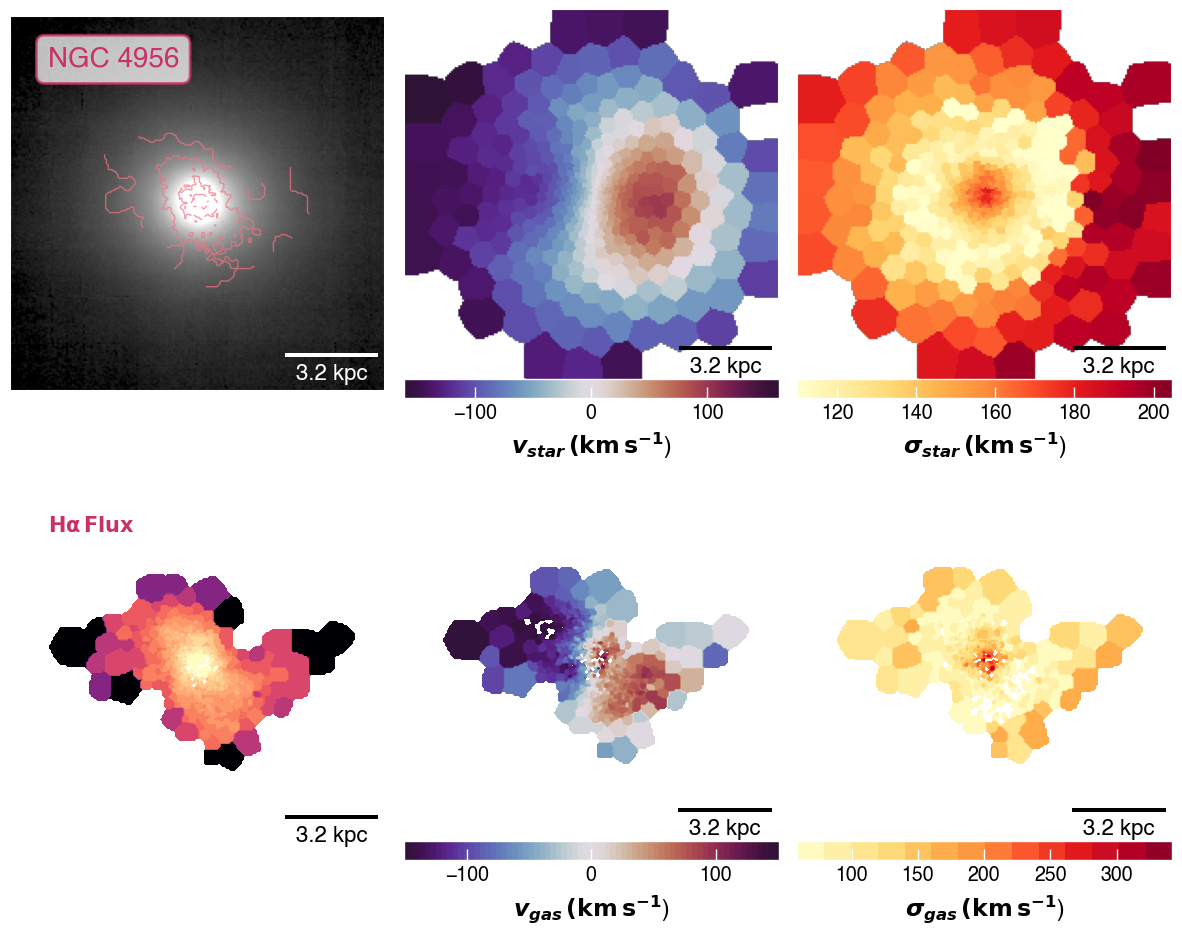}

    \caption{Same as Fig.\ref{fig:gas_distribution}}
    \label{fig:Halpha_distribution6}
\end{figure*}

\begin{figure*}[ht!]    
    \centering
    \includegraphics[height=0.65\textwidth]{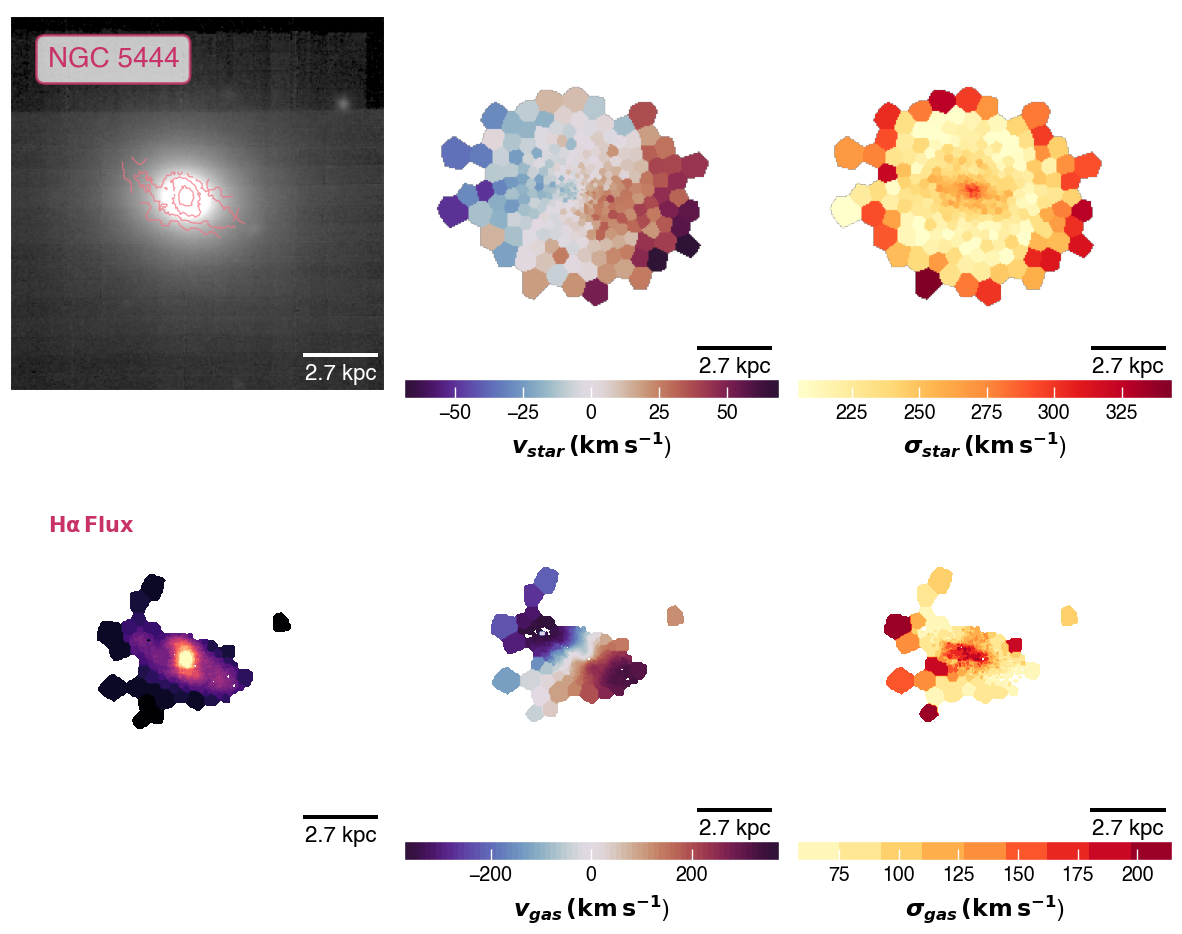}
    \caption{Same as Fig.\ref{fig:gas_distribution}}
    \label{fig:Halpha_distribution7}
\end{figure*}

\end{appendix}

\end{document}